\documentclass[11pt]{article}

\usepackage[letterpaper,margin=1in]{geometry}
\usepackage{amsmath,amssymb,amsthm,mathtools}
\usepackage{booktabs,array,longtable,tabularx}
\usepackage{float}
\usepackage[section]{placeins}
\usepackage{graphicx}
\usepackage{xurl}
\usepackage{enumitem}
\usepackage{microtype}
\usepackage{xcolor}
\usepackage{caption}
\usepackage{titlesec}
\usepackage{fancyhdr}
\usepackage[hidelinks]{hyperref}

\definecolor{accent}{RGB}{35,72,110}
\hypersetup{
  colorlinks=true,
  linkcolor=accent,
  urlcolor=accent,
  citecolor=accent,
  pdftitle={A Pair-Weighted Reciprocal-Cost Representation of the von Weizsaecker Functional},
  pdfauthor={Megan Simons and Jonathan Washburn},
  pdfkeywords={von Weizsaecker functional, orbital-free density-functional theory, Fisher information, reciprocal cost, Dirichlet energy},
}

\titleformat{\section}{\large\bfseries\color{accent}}{\thesection}{0.7em}{}
\titleformat{\subsection}{\normalsize\bfseries}{\thesubsection}{0.7em}{}
\titleformat{\subsubsection}{\normalsize\itshape}{\thesubsubsection}{0.7em}{}

\setlist[itemize]{leftmargin=1.5em,itemsep=0.15em,topsep=0.3em}
\setlist[enumerate]{leftmargin=1.7em,itemsep=0.15em,topsep=0.3em}

\theoremstyle{plain}
\newtheorem{theorem}{Theorem}[section]
\newtheorem{proposition}[theorem]{Proposition}
\newtheorem{lemma}[theorem]{Lemma}
\newtheorem{corollary}[theorem]{Corollary}
\theoremstyle{definition}
\newtheorem{definition}[theorem]{Definition}
\newtheorem{assumption}[theorem]{Assumption}
\theoremstyle{remark}
\newtheorem{remark}[theorem]{Remark}

\newcommand{\TW}{T_{\mathrm{W}}}
\newcommand{\Ts}{T_{\mathrm{s}}}
\newcommand{\Tth}{T_{\theta}}
\newcommand{\Tdet}{T_{\mathrm{det}}}
\newcommand{\TPdet}{T_{\mathrm{P,det}}}
\newcommand{\fWdet}{f_{\mathrm{W}}^{\mathrm{det}}}
\newcommand{\IF}{I_{\mathrm{F}}}
\newcommand{\rr}{\mathbf{r}}
\newcommand{\Ecal}{\mathcal{E}}
\newcommand{\Scal}{\mathcal{S}}
\newcommand{\Kcal}{\mathcal{K}}
\newcommand{\dr}{\,d\rr}
\newcommand{\Ha}{\mathrm{Ha}}

\begin{document}

\begin{center}
{\LARGE\bfseries A Pair-Weighted Reciprocal-Cost Representation\\[0.25em]
of the von Weizs\"acker Functional}\\[1.6em]
{\large Megan Simons\footnote{Corresponding author:
\texttt{msimons@recognitionphysics.org}}\quad and\quad Jonathan Washburn}\\[0.5em]
{\normalsize\itshape Recognition Science, Recognition Physics Institute,
Austin, Texas, USA}
\end{center}

\vspace{0.6em}
\hrule
\vspace{1.0em}

\begin{abstract}
\noindent
We investigate pair-weighted lattice energies constructed from the reciprocal
ratio cost $J(x)=\tfrac12(x+x^{-1})-1$ and their connection to the
von Weizs\"acker kinetic-energy functional. When the pair weight depends
only on the common multiplicative amplitude scale and is quadratic under
uniform amplitude rescaling, it takes the form $B(u,v)=Cuv$. For the
normalization $C=1$, the weighted edge cost reduces exactly to
$\tfrac12(u-v)^2$ for arbitrary positive neighboring amplitudes, without
requiring a small-deviation expansion. On a Cartesian lattice, the resulting
discrete energy converges to $\tfrac12\int|\nabla u|^2\dr$, with leading
correction
$-\tfrac{h^2}{24}\sum_k\int(\partial_k^2 u)^2\dr+O(h^4)$
under the stated regularity conditions. Identifying $u=\sqrt{\rho}$ and
restoring the factor $\hbar^2/m$ yields the von Weizs\"acker functional.
By contrast, the unweighted ratio cost approaches a scale-invariant
log-Dirichlet action and does not have kinetic-energy dimensions. Numerical
benchmarks confirm the expected lattice convergence and assess the
discretization error in smooth, nodal, cusp, and molecular cases. For
H$_2^+$ densities obtained from aug-cc-pV5Z orbitals, a four-grid
least-squares $h^2+h^4$ extrapolation agrees with orbital kinetic energies
within $1.5\times10^{-4}$ relative over $R=1$--$6\,a_0$. Alternative
extrapolations indicate a conservative sensitivity of up to approximately
$5\times10^{-4}$. The construction therefore provides a conditional exact
discrete representation of the von Weizs\"acker sector rather than an
approximation to the complete noninteracting kinetic functional.
\end{abstract}

\vspace{0.8em}
\hrule
\vspace{0.6em}


\section{Introduction}
\label{sec:intro}

Density-functional theory (DFT) asserts that the ground-state energy is a
functional of the electron density alone~\cite{HohenbergKohn1964}. In practice
the theory is almost always implemented in Kohn--Sham form~\cite{KohnSham1965},
in which orbitals are reintroduced primarily to evaluate the noninteracting
kinetic energy $\Ts[\rho]$ accurately. No explicit, generally accurate density
functional for $\Ts$ is known. An accurate and transferable approximation to
$\Ts[\rho]$ is therefore central to practical orbital-free
DFT~\cite{WangCarter2000,KarasievTrickey2012,Witt2018,Mi2023,Xu2024}. The
oldest explicit candidate is the Thomas--Fermi
term~\cite{Thomas1927,Fermi1927}, local in $\rho$ and exact for the uniform
electron gas. Gradient corrections are systematically defined for slowly
varying densities~\cite{Kirzhnits1957,Hodges1973} but can be inaccurate in
molecular bonding and surface
regions~\cite{WangCarter2000,KarasievTrickey2012,Witt2018}.

Let $\rho(\rr)\ge0$ denote an electron density, normalized by
$\int\rho(\rr)\,d\rr=N$ for an $N$-electron system, and let $\rr$ denote the
spatial coordinate. An exact lower-bound contribution is the von Weizs\"acker
functional~\cite{Weizsaecker1935},
\begin{equation}
\label{eq:TW}
\TW[\rho]
=\frac{\hbar^2}{8m}\int\frac{|\nabla\rho(\rr)|^2}{\rho(\rr)}\dr
=\frac{\hbar^2}{2m}\int\bigl|\nabla\sqrt{\rho(\rr)}\bigr|^2\dr ,
\end{equation}
where $\hbar$ is the reduced Planck constant and $m$ is the electron mass. For
nonnegative densities, the second expression defines $\TW$ on the natural
domain $\sqrt\rho\in H^1$, the Sobolev space of square-integrable functions
with square-integrable weak first derivatives. The quotient form is equivalent
for positive, sufficiently regular densities and is understood through this
definition at zeros. The functional is the exact kinetic energy of any density
that comes from a single real orbital. It is a rigorous lower bound to $\Ts$
for admissible $N$-representable
densities~\cite{HoffmannOstenhof1977,Lieb1983}. In the slowly varying
density-gradient expansion, one ninth of $\TW$ appears as the second-order
gradient correction~\cite{Kirzhnits1957,Hodges1973}. Up to a constant it is
the Fisher information of the
density~\cite{Fisher1925,SearsParrDinur1980,Nagy2003,Nagy2025}. For general
many-orbital densities, however, $\TW[\rho]\ne\Ts[\rho]$. The remainder
\begin{equation}
\label{eq:split}
\Ts[\rho]=\TW[\rho]+\Tth[\rho],
\qquad \Tth[\rho]\ge0,
\end{equation}
is the Pauli term~\cite{LevyPerdewSahni1984,LevyOuYang1988,Herring1986}, and
$\Tth$ typically carries a substantial fraction of the kinetic energy.
Approximating and differentiating this sector remain active orbital-free-DFT
problems~\cite{MajumdarShiVignale2025}.

Sears, Parr and Dinur related $\TW$ to information content in the
density~\cite{SearsParrDinur1980}, and Reginatto obtained the Schr\"odinger
equation from a principle of minimum Fisher
information~\cite{Reginatto1998}. Those arguments take the Dirichlet energy in
$\sqrt\rho$ as input. We start from a scalar cost on positive ratios,
characterized independently of densities and kinetic
energy~\cite{PardoGuerra2026,WashburnZlatanovic2026}, and ask what continuum
functional results under the pair-weight assumptions introduced below.
The cost is
\begin{equation}
\label{eq:J}
J:(0,\infty)\to[0,\infty),
\qquad
J(x)=\frac12\left(x+\frac1x\right)-1 .
\end{equation}

The square-root expression $(\sqrt{\rho_i}-\sqrt{\rho_j})^2$ is already a
discrete Fisher/Dirichlet quantity. For a probability vector on a path, Chen,
Liu and Saloff-Coste define a discrete Fisher information
$2\sum_i\bigl|\sqrt{p_i}-\sqrt{p_{i-1}}\bigr|^2$ and show that it equals a
multiple of the Dirichlet form of
$\sqrt{p}$~\cite[p.~246]{ChenLiuSaloffCoste2008}. The square-root
Dirichlet/Fisher structure is already established. Here we factor the
independently characterized ratio cost into that form under the stated
pair-weight assumptions and derive the leading lattice error and node
extension. Real-space finite-difference discretizations of orbital-free DFT,
including the von Weizs\"acker term, are already
established~\cite{MiATLAS2016}; the finite-difference form itself is not
claimed as new. The main contribution is to connect the independently
characterized reciprocal cost $J$ to the established square-root Dirichlet
form under explicit pair-weight assumptions. Within the
scale--deviation-separated, degree-two class, the pair weight is $B=Cuv$;
with $C=1$, the resulting edge cost is exactly $\tfrac12(u-v)^2$ at every
positive ratio. We also analyze the continuous node extension and Cartesian
lattice error. Conversely, within the geometric-mean pair-weight class, a
nonzero nonnegative quadratic edge cost that vanishes on the diagonal forces
$F\propto J$ (Appendix~\ref{app:characterizations}).

For amplitudes $u,v>0$, the coordinates $s=\sqrt{uv}$ and $r=v/u$ separate
the common scale from the dimensionless mismatch. Restricting the pair weight
to the common scale, together with degree-two amplitude covariance, gives
$B(u,v)=Cuv$ (Proposition~\ref{prop:scaleonly}). With the calibration $C=1$,
\begin{equation}
\label{eq:main}
uv\,J\!\left(\frac{v}{u}\right)=\frac12(u-v)^2
\end{equation}
for every $u,v>0$, with no small-deviation expansion
(Proposition~\ref{prop:identity}). On a Cartesian lattice the corresponding
edge energy converges to $\tfrac12\int|\nabla u|^2\dr$, with a negative
leading $h^2$ correction under the hypotheses of
Theorem~\ref{thm:continuum}. Setting $u=\sqrt{\rho}$ and restoring
$\hbar^2/m$ gives $\TW$.
Scale--deviation separation is a modeling assumption, while $C=1$,
$u=\sqrt{\rho}$, and $\hbar^2/m$ are fixed separately; the resulting
construction recovers the von Weizs\"acker sector rather than the full
$\Ts[\rho]$.
The same ratio cost without
the pair weight converges to a scale-invariant log-Dirichlet action of
dimensions $L^{d-2}$ (Section~\ref{sec:unbridged}) and does not have
kinetic-energy dimensions.

Atomic units ($\hbar=m=e=1$) are used in all numerical work; factors of
$\hbar^2/m$ are displayed in the formal statements wherever the distinction
between functional shape and physical scale matters. We denote the Bohr
radius by $a_0$ and the Hartree energy unit by $\Ha$. Throughout, $\log$ is the
natural logarithm, and $\varepsilon_{\text{mach}}=2^{-52}\approx2.22\times
10^{-16}$ is the double-precision machine epsilon of the arithmetic used in
Section~\ref{sec:results}.

\section{Reciprocal cost and the von Weizs\"acker decomposition}
\label{sec:setting}

\subsection{The function \texorpdfstring{$J$}{J} and the composition law}

Throughout, $J$ is the function \eqref{eq:J}.

\begin{lemma}
\label{lem:J}
For $x>0$: (i) $J(x)\ge0$ with equality iff $x=1$; (ii) $J(x)=J(1/x)$;
(iii) $J$ is strictly convex on $(0,\infty)$; (iv) in logarithmic coordinates
$J(e^{\epsilon})=\cosh\epsilon-1$.
\end{lemma}

\begin{proof}
$J(x)=\frac{(x-1)^2}{2x}$, which gives (i) immediately. (ii) is invariance of
$x+x^{-1}$ under $x\mapsto x^{-1}$. (iii) $J''(x)=x^{-3}>0$. (iv) is direct
substitution.
\end{proof}

The factorization
\begin{equation}
\label{eq:Jfact}
J(x)=\frac{(x-1)^2}{2x}
\end{equation}
writes $J$ as a squared difference divided by the ratio. Multiplication by
$uv$ cancels that denominator. In logarithmic coordinates,
\begin{equation}
\label{eq:Jcosh}
J(e^{\epsilon})=\cosh\epsilon-1=\frac{\epsilon^2}{2}+\frac{\epsilon^4}{24}+O(\epsilon^6).
\end{equation}

Washburn and Zlatanovi\'c~\cite{WashburnZlatanovic2026} prove that a
solution of the composition law
\begin{equation}
\label{eq:rcl}
F(xy)+F(x/y)=2F(x)F(y)+2F(x)+2F(y),
\end{equation}
satisfying the unit logarithmic-curvature calibration
$\lim_{t\to0}2F(e^t)/t^2=1$, is uniquely $J$. That characterization does not
use gradients, densities, or kinetic energy.
Appendix~\ref{app:characterizations} gives a self-contained proof
(Theorem~\ref{thm:Juniqueness}) in which the curvature condition itself
implies continuity. For continuous nonconstant $F$ with $F(1)=0$
and a symmetric polynomial combiner of degree at most two, the combiner
reduces to the d'Alembert family
$2u+2v+cuv$~\cite{WashburnZlatanovicAllahyarov2026}. The d'Alembert
classification of that equation is classical~\cite{Aczel1966}. A broader
classification of admissible reciprocally symmetric costs is given in
Ref.~\cite{PardoGuerraWashburnAllahyarov2026}.
Previous work has studied $J$ through composition laws and
reciprocal-symmetry classifications; here $J$ is instead applied to spatial
nearest-neighbor amplitudes.

\subsection{von Weizs\"acker energy, Fisher information, and the Pauli term}

For a positive density $\rho$ with $u=\sqrt\rho$, one has $\nabla\rho=2u\nabla
u$ and hence $|\nabla\rho|^2/\rho=4|\nabla u|^2$, so the two forms in
\eqref{eq:TW} agree. Writing the Fisher information of the
density~\cite{Fisher1925} as
\begin{equation}
\label{eq:fisher}
\IF[\rho]=\int\frac{|\nabla\rho|^2}{\rho}\dr,
\end{equation}
the von Weizs\"acker energy is $\TW[\rho]=\tfrac{\hbar^2}{8m}\IF[\rho]$: it is
the unnormalized density Fisher functional in energy
units~\cite{SearsParrDinur1980,Nagy2003,Nagy2025}. For the probability density
$p=\rho/N$, the conventional Fisher information satisfies
$I[p]=\IF[\rho]/N$.
Reginatto's derivation of the Schr\"odinger equation from minimum Fisher
information~\cite{Reginatto1998} starts from this identification. Here the
same functional arises after $J$ is paired with the specified weight and the
amplitude is identified with $\sqrt{\rho}$.

For any $N$-representable density, \eqref{eq:split} determines the
nonnegative Pauli term
$\Tth$~\cite{LevyPerdewSahni1984,LevyOuYang1988,Herring1986}. A single
occupied spatial orbital implies $\Tth=0$; a converse requires the usual
connected-support and regularity conditions excluding disjoint orbital
supports. Section~\ref{sec:scope} shows explicitly why the amplitude-only
construction recovers $\TW$ but not the current or Pauli contributions to
the full kinetic energy.

\section{Pair weights}
\label{sec:bridge}

\begin{definition}[State-dependent pair weight]
\label{def:bridge}
A \emph{state-dependent pair weight} is a function
$B:(0,\infty)^2\to(0,\infty)$ used to weight the reciprocal cost of a
neighboring pair. The edge cost is
\begin{equation}
\label{eq:edgecost}
\Kcal(u,v)=B(u,v)\,F\!\left(\frac{v}{u}\right)
\end{equation}
for a reciprocal-symmetric cost $F$, that is $F(x)=F(1/x)$.
\end{definition}

We impose two structural requirements on $B$:

\begin{assumption}[Symmetry and degree-two covariance]
\label{ass:bridge}
\leavevmode
\begin{enumerate}[label=(B\arabic*)]
\item Symmetry: $B(u,v)=B(v,u)$. An edge has no preferred orientation.
\item Degree-two covariance: $B(cu,cv)=c^2B(u,v)$ for all $c>0$. A
      uniform rescaling of the amplitude rescales the pair weight as the
      squared amplitude, matching the degree-two scaling of the normalization
      $\int u^2$.
\end{enumerate}
\end{assumption}

\subsection{Symmetric homogeneous weights}

Conditions (B1)--(B2) leave a residual ratio-dependent freedom.

\begin{lemma}
\label{lem:structure}
A function $B$ satisfies \textup{(B1)} and \textup{(B2)} if and only if it can
be written
\begin{equation}
\label{eq:Bform}
B(u,v)=uv\,\chi\!\left(\frac{v}{u}\right)
\qquad\text{with}\qquad
\chi:(0,\infty)\to(0,\infty),
\quad
\chi(x)=\chi(1/x).
\end{equation}
\end{lemma}

\begin{proof}
Any degree-two homogeneous $B$ may be written $B(u,v)=u^2b(v/u)$ with
$b(x)=B(1,x)$. Substituting $b(x)=x\chi(x)$ gives $B(u,v)=uv\,\chi(v/u)$,
which is legitimate since $x>0$. Symmetry then reads $uv\chi(v/u)
=uv\chi(u/v)$, i.e.\ $\chi(x)=\chi(1/x)$. Conversely any $\chi$ with
that invariance yields a $B$ obeying (B1) and (B2).
\end{proof}

Symmetry and degree-two covariance do not force $B\propto uv$.
The choice $\chi(x)=x/(1+x)^2$ satisfies $\chi(x)=\chi(1/x)$ and
gives $B(u,v)=u^{2}v^{2}/(u+v)^{2}$. This weight is symmetric, degree-two
homogeneous and strictly positive, and it is not a multiple of $uv$.

An additional condition removes this freedom: separation of the pair scale
from the relative deviation.

\subsection{Scale--deviation-separated weights}

Because the reciprocal cost is naturally expressed in logarithmic
coordinates, we separate a pair into its logarithmic midpoint and
difference,
\begin{equation}
\label{eq:log-midpoint}
\sigma=\tfrac12\bigl(\log(u/u_\ast)+\log(v/u_\ast)\bigr),
\qquad
\delta=\log(v/u),
\end{equation}
where $u_\ast>0$ is an arbitrary fixed reference amplitude.
Exponentiation gives $u_\ast e^{\sigma}=\sqrt{uv}$ and $e^{\delta}=v/u$.
Thus $\sqrt{uv}$ is the multiplicative midpoint associated with the ratio
coordinate used by $J$. Equivalently, in amplitude variables,
\begin{equation}
\label{eq:scale-ratio}
s=\sqrt{uv},
\qquad
r=\frac vu,
\qquad
u=\frac{s}{\sqrt r},
\quad
v=s\sqrt r .
\end{equation}
Here $s$ is the common amplitude scale and $r$ is the departure from
equilibrium.

The factorization separates two independent attributes of a pair: its common
amplitude scale and its dimensionless mismatch. Since $F(v/u)$ is already
defined as the cost of relative deviation, assigning additional ratio
dependence to the pair weight would place relative-deviation dependence in
both factors. In the separable class, that dependence is carried entirely by
$F$. That assignment is a modeling choice, not a consequence of symmetry or
covariance, and it is needed for the exact finite-ratio form rather than for
the leading continuum limit.

\begin{assumption}[Scale--deviation separation]
\label{ass:scale-deviation}
The ratio cost $F(r)$ carries the complete dependence of the edge cost on
relative deviation. The pair weight carries only the pair scale:
\begin{equation}
\label{eq:scale-separation}
B(u,v)=b(s)=b\!\left(\sqrt{uv}\right)
\end{equation}
for some $b:(0,\infty)\to(0,\infty)$.
\end{assumption}

\begin{proposition}
\label{prop:scaleonly}
Suppose $B$ obeys \textup{(B2)} and
Assumption~\ref{ass:scale-deviation}. Then
$B(u,v)=C\,uv$ with $C=b(1)>0$. No regularity assumption on $b$ is required.
\end{proposition}

\begin{proof}
Under $(u,v)\mapsto(cu,cv)$, the scale coordinate transforms as $s\mapsto cs$.
Degree-two covariance therefore gives $b(cs)=c^2b(s)$ for all $c,s>0$.
Setting $s=1$ yields $b(c)=c^2b(1)$, so
$B(u,v)=b(\sqrt{uv})=b(1)uv$.
\end{proof}

\begin{remark}
No continuity or measurability assumption on $b$ is needed because the
homogeneity relation holds for every $c>0$. For amplitudes $u_i=\sqrt{\rho_i}$,
$uv=\sqrt{\rho_i\rho_j}$ is the geometric mean of the neighboring densities,
and $B=Cuv$ is the geometric-mean pair weight.
\end{remark}

\begin{remark}
\label{rem:inversion}
The geometric mean can also be characterized by an inversion property.
If a symmetric degree-one homogeneous pair scale satisfies
$s(u_\ast^2/u,u_\ast^2/v)=u_\ast^2/s(u,v)$, then $s=\sqrt{uv}$.
The analogous degree-two result gives $B\propto uv$
(Proposition~\ref{prop:inversion}).
\end{remark}

The dimensionless edge normalization is set to $C=1$. A general
$C$ would multiply the pair-weighted cost, every lattice functional
derived from it, and its continuum limit by $C$; matching the conventional normalization
of $\TW$ therefore requires $C=1$. The separate dimensional kinetic scale is
restored in Section~\ref{sec:identification}.

\subsection{The unweighted ratio cost}
\label{sec:unbridged}

Neighboring ratios alone define the sum
\begin{equation}
\label{eq:logaction}
\sum_{\langle ij\rangle}J\!\left(\frac{u_j}{u_i}\right)
=\sum_{\langle ij\rangle}\bigl[\cosh(\phi_i-\phi_j)-1\bigr],
\qquad \phi_i=\log(u_i/u_\ast),
\end{equation}
where $u_\ast>0$ is an arbitrary fixed reference amplitude. Differences
satisfy $\phi_j-\phi_i=\log(u_j/u_i)$, so $u_\ast$ cancels.
For small $\phi_i-\phi_j$ the summand is $\tfrac12(\phi_i-\phi_j)^2$. Any
smooth reciprocal-symmetric cost $F(x)=F(1/x)$ with $F(1)=0$ and the same
unit logarithmic curvature has the same leading quadratic term, so the
expansion does not select the reciprocal form of $J$. The same quadratic expansion is used in discrete
models based on this cost~\cite{PardoGuerra2026}. With the lattice weight
$h^{d-2}$, the continuum limit is the Dirichlet energy of $\log(u/u_\ast)$.

\begin{proposition}
\label{prop:unbridged}
Let $\Omega$ be a bounded rectangular domain and let
$u\in C^{2}(\overline{\Omega})$ satisfy $u\ge u_{\min}>0$ on
$\overline{\Omega}$. Define the unweighted
lattice cost over the Cartesian edges contained in $\Omega$ by
\begin{equation}
\Scal_h^{\mathrm{unw}}[u]
=h^{d-2}\sum_{(i,j)\in E_h}J\!\left(\frac{u_j}{u_i}\right).
\end{equation}
Then
\begin{equation}
\label{eq:unbridged}
\Scal_h^{\mathrm{unw}}[u]
\;\xrightarrow[h\to0]{}\;
\Scal^{\mathrm{unw}}[u]
\equiv
\frac12\int\bigl|\nabla\log u\bigr|^{2}\dr
=\frac18\int\frac{|\nabla\rho|^{2}}{\rho^{2}}\dr ,
\qquad u=\sqrt\rho .
\end{equation}
This action is invariant under $\rho\mapsto c\rho$ for $c>0$ and has dimensions
$L^{d-2}$. It is not an energy: multiplying it by $\hbar^2/m$ would give
energy times volume rather than energy.
\end{proposition}

\begin{proof}
The lower bound $u\ge u_{\min}>0$ on the closed domain implies
$\phi=\log(u/u_\ast)\in C^{2}(\overline{\Omega})$, so
$J(u_j/u_i)=\cosh(\phi_j-\phi_i)-1$
and $\nabla\phi=\nabla\log u$. For a Cartesian edge in direction $k$,
$\phi_j-\phi_i=h\,\partial_k\phi
+O(h^{2})$, hence $\cosh(\phi_j-\phi_i)-1=\tfrac12h^{2}(\partial_k\phi)^{2}
+O(h^{3})$. Summing with the weight $h^{d-2}$ converts the edge sum into a
Riemann sum with cell volume $h^{d}$, giving $\tfrac12\int|\nabla\phi|^2$.
With $u=\sqrt\rho$ one has $\nabla\log u=\tfrac12\nabla\rho/\rho$, which gives
\eqref{eq:unbridged}. Scale invariance is immediate since $\nabla\log(c\rho)
=\nabla\log\rho$.
\end{proof}

As a function on $\mathbb{R}^3$, the hydrogen $1s$ amplitude is not
differentiable at the nucleus and therefore is not $C^2$; it lies
outside the proposition's hypotheses. Direct evaluation away from that
measure-zero point nevertheless gives
$\nabla\log\sqrt\rho=-\hat{\mathbf r}/a_0$ and
$|\nabla\log\sqrt\rho|^2=a_0^{-2}$ almost everywhere. Its unweighted continuum
action equals $|\Omega|/(2a_0^2)$, with dimensions of length in three
dimensions, and diverges with the domain volume. In atomic units $a_0=1$, so
the numerical value reduces to $\tfrac12|\Omega|$.

\begin{remark}
\label{rem:bridge-supplies}
Comparing \eqref{eq:unbridged} with Corollary~\ref{cor:TW},
\begin{equation}
\underbrace{\frac18\int\frac{|\nabla\rho|^{2}}{\rho^{2}}\dr}
_{\text{unweighted action }[L^{d-2}]}
\qquad\text{versus}\qquad
\underbrace{\frac18\int\frac{|\nabla\rho|^{2}}{\rho}\dr}
_{\text{pair-weighted form }[L^{-2}]},
\end{equation}
the product of neighboring amplitudes tends to the local density,
$u_iu_j=\sqrt{\rho_i\rho_j}\to\rho$. Multiplication by $u_iu_j$ changes the
density scaling from degree zero to degree one and changes the continuum
dimensions from $L^{d-2}$ to $L^{-2}$. The limit then acquires kinetic-energy
units after multiplication by $\hbar^2/m$.
Section~\ref{sec:res-identity} checks the volume divergence numerically.
\end{remark}

\section{The pair cost as a Dirichlet edge energy}
\label{sec:identity}

\subsection{Exact identity on the positive quadrant}

\begin{proposition}
\label{prop:identity}
Define $\Kcal_J(u,v)=u\,v\,J(v/u)$ for $u,v>0$. Then
\begin{equation}
\label{eq:prop1}
\Kcal_J(u,v)=\frac12\,(u-v)^2
\qquad\text{for all }u,v>0 .
\end{equation}
\end{proposition}

\begin{proof}
For $u,v>0$,
\begin{align*}
uv\,J\!\left(\frac{v}{u}\right)
&=uv\left[\frac12\left(\frac{v}{u}+\frac{u}{v}\right)-1\right]
=\frac12(u^2+v^2)-uv
=\frac12(u-v)^2 .
\end{align*}
\end{proof}

\begin{remark}
Equation~\eqref{eq:prop1} holds for arbitrary positive $v/u$.
Section~\ref{sec:res-identity} tests the floating-point evaluation of the
same identity near $v/u=1$.
\end{remark}

\subsection{Continuous extension to vanishing amplitudes}

The ratio $v/u$ in \eqref{eq:prop1} is undefined when $u=0$, but quantum
amplitudes have nodes, and densities vanish at the boundaries of finite
systems. The right-hand side has no such difficulty, and it gives the only
continuous way to remove the singularity.

\begin{proposition}
\label{prop:nodes}
Let $Q=[0,\infty)^2$ and $Q^{+}=(0,\infty)^2$. There is exactly one continuous
function $\overline{\Kcal}_J:Q\to[0,\infty)$ agreeing with $\Kcal_J$ on $Q^{+}$,
namely
\begin{equation}
\label{eq:ext}
\overline{\Kcal}_J(u,v)=\frac12\,(u-v)^2 ,
\qquad u,v\ge0 .
\end{equation}
In particular $\overline{\Kcal}_J(0,v)=\tfrac12v^2$,
$\overline{\Kcal}_J(u,0)=\tfrac12u^2$ and $\overline{\Kcal}_J(0,0)=0$.
\end{proposition}

\begin{proof}
Existence: $(u,v)\mapsto\tfrac12(u-v)^2$ is continuous on $Q$ and agrees with
$\Kcal_J$ on $Q^{+}$ by Proposition~\ref{prop:identity}.

Uniqueness: $Q^{+}$ is dense in $Q$, since any $(u,v)\in Q$ is the limit of
$(u+1/n,\,v+1/n)\in Q^{+}$. Two continuous functions on $Q$ that agree on a
dense subset agree everywhere.
\end{proof}

\begin{remark}
The ratio representation is undefined at a node, whereas the extension assigns
the finite cost $\tfrac12v^2$ to an edge joining a node to an amplitude $v$.
The particle-in-a-box benchmark of Section~\ref{sec:res-convergence} converges
at second order through the nodes at both walls. The extension
removes the discrete ratio singularity only: continuum convergence still
requires the regularity in Theorem~\ref{thm:continuum}. In particular, the box
benchmark tests boundary nodes and does not establish second-order convergence
across arbitrary interior nodal surfaces, where $\sqrt\rho$ may be nonsmooth.
\end{remark}

\section{Lattice Dirichlet energy}
\label{sec:lattice}

\subsection{Nearest-neighbor energy on a Cartesian lattice}

Let $\Omega\subseteq\mathbb{R}^d$ and let $G_h$ be the regular Cartesian
lattice of spacing $h$ on $\Omega$, with vertex set $V_h$ and nearest-neighbor
edge set $E_h$, each edge counted once. Assign $u_i\ge0$ to each vertex.

\begin{definition}[Pair-weighted lattice energy]
\label{def:Eh}
For a strictly positive vertex field, $u_i>0$ for every vertex $i$, define
\begin{equation}
\label{eq:Eh}
\Ecal_h[u]=h^{d-2}\sum_{(i,j)\in E_h}u_iu_j\,J\!\left(\frac{u_j}{u_i}\right).
\end{equation}
For a nonnegative field, define each edge term by the continuous extension
$\overline{\Kcal}_J(u_i,u_j)=\tfrac12(u_i-u_j)^2$ of
Proposition~\ref{prop:nodes}.
\end{definition}

Dimensional analysis gives the prefactor $h^{d-2}$: a nearest-neighbor
difference scales as $h$, its square as $h^2$, and a Riemann sum over lattice
cells carries $h^d$. Under this nearest-neighbor scaling, the required
prefactor is $h^{d-2}=h^{d}/h^{2}$, which produces a finite, nonzero
continuum Dirichlet integral.

\begin{proposition}
\label{prop:lattice}
For any $u:V_h\to[0,\infty)$,
\begin{equation}
\label{eq:latticeeq}
\Ecal_h[u]=\frac{h^{d-2}}{2}\sum_{(i,j)\in E_h}(u_i-u_j)^2 .
\end{equation}
More generally, on any finite weighted graph $G=(V,E,w)$ with $w_{ij}>0$,
\begin{equation}
\label{eq:grapheq}
\sum_{(i,j)\in E}w_{ij}\,u_iu_j\,J\!\left(\frac{u_j}{u_i}\right)
=\frac12\sum_{(i,j)\in E}w_{ij}(u_i-u_j)^2 ,
\end{equation}
the standard weighted graph Dirichlet form, with the left-hand side
understood in this continuously extended sense when either endpoint
amplitude vanishes.
\end{proposition}

\begin{proof}
Apply Proposition~\ref{prop:identity} edgewise and sum; the extension to
vanishing amplitudes is Proposition~\ref{prop:nodes}.
\end{proof}

Equation~\eqref{eq:grapheq} holds on arbitrary graphs, with no lattice
structure, no notion of dimension and no smallness assumption. The regular
lattice is needed only for the continuum limit, not for the equivalence.

\subsection{Variational structure and the lattice Laplacian}
\label{sec:variational}

Lattice fields are compared to continuum functions through the discrete $L^2$
inner product $\langle a,b\rangle_h=h^{d}\sum_ia_ib_i$, the Riemann-sum
approximation to $\int ab\dr$. The gradient of a lattice functional with
respect to this inner product is
\begin{equation}
\label{eq:disc-grad-def}
(\nabla_h\Ecal_h)_i=h^{-d}\,\frac{\partial\Ecal_h}{\partial u_i},
\end{equation}
the factor $h^{-d}$ converting a derivative with respect to a vertex value
into a density per unit volume. With this normalization the gradient of the
lattice energy is the negative lattice Laplacian.

\begin{proposition}
\label{prop:gradient}
For an interior vertex $i$ of $G_h$,
\begin{equation}
\label{eq:grad}
\frac{\partial\Ecal_h}{\partial u_i}
=h^{d-2}\sum_{j\sim i}(u_i-u_j),
\end{equation}
and therefore
\begin{equation}
\label{eq:latlap}
(\nabla_h\Ecal_h)_i
=\frac{1}{h^{2}}\sum_{j\sim i}(u_i-u_j)
=-\Delta_h u_i ,
\end{equation}
where $\Delta_h$ is the standard $(2d+1)$-point lattice Laplacian.
Consequently, at interior lattice points,
\begin{equation}
\label{eq:latlap-pt}
(\nabla_h\Ecal_h)_i=-\nabla^2u(\rr_i)+O(h^2)
\end{equation}
for $u\in C^{4}$, with the error uniform on compact interior subsets when
the fourth derivatives are bounded.
\end{proposition}

\begin{proof}
Vertex $i$ appears in exactly the edges $(i,j)$ with $j\sim i$, each
contributing $\tfrac{h^{d-2}}{2}(u_i-u_j)^2$ by \eqref{eq:latticeeq};
differentiating gives \eqref{eq:grad}. Dividing by $h^{d}$ as in
\eqref{eq:disc-grad-def} gives the first equality in \eqref{eq:latlap}. The
standard central-difference identity in direction $k$ reads
$u(\rr+h\mathbf{e}_k)+u(\rr-h\mathbf{e}_k)-2u(\rr)=h^2\partial_k^2u+O(h^4)$,
so $h^{-2}\sum_{j\sim i}(u_i-u_j)=-\sum_k\partial_k^2u+O(h^2)=-\nabla^2u+O(h^2)$.
If the fourth derivatives of $u$ are bounded on a compact interior set, the
$O(h^4)$ remainder is uniform there, and so is the $O(h^2)$ in
\eqref{eq:latlap-pt}.
\end{proof}

The expansion \eqref{eq:Jcosh} produces this quadratic edge form to leading
order near $x=1$. Proposition~\ref{prop:lattice} establishes the form for
every positive ratio, before the continuum approximation of
Proposition~\ref{prop:gradient} and Theorem~\ref{thm:continuum}.

\section{Lattice-to-continuum limit}
\label{sec:continuum}

Under the assumptions below, the leading discretization error can be
evaluated explicitly. The coefficient is available in closed form for the
analytic benchmarks and is compared with the measured discretization error in
Section~\ref{sec:res-convergence}.

\begin{theorem}
\label{thm:continuum}
Let either (i) $u\in\mathcal{S}(\mathbb{R}^d)$, the Schwartz space of smooth
rapidly decreasing functions, be sampled on a full Cartesian lattice of
spacing $h$ (with any fixed shift), or (ii) $u\in C^5(\mathbb{T}^d)$ be sampled
on a periodic Cartesian grid commensurate with the periods of the
$d$-dimensional torus $\mathbb{T}^d$. Assume $u\ge0$ and sum over all
nearest-neighbor edges.
Then
\begin{equation}
\label{eq:continuum}
\Ecal_h[u]
=\frac12\int|\nabla u|^2\dr
\;-\;\frac{h^{2}}{24}\,A[u]
\;+\;O(h^{4}),
\qquad
A[u]=\sum_{k=1}^{d}\int\bigl(\partial_k^{2}u\bigr)^{2}\dr .
\end{equation}
In particular $\Ecal_h[u]\to\tfrac12\int|\nabla u|^2\dr$ with second-order
accuracy. If $A[u]>0$, the leading error is negative, so the approach is from
below for sufficiently small $h$.
\end{theorem}

\begin{proof}
By Proposition~\ref{prop:lattice}, $\Ecal_h[u]=\tfrac{h^{d-2}}{2}
\sum_{(i,j)}(u_i-u_j)^2$. Group the edges by direction $k$ and parametrize each
by its midpoint $\mathbf{m}$. A symmetric Taylor expansion about the midpoint
gives
\begin{equation}
u(\mathbf{m}+\tfrac{h}{2}\mathbf{e}_k)-u(\mathbf{m}-\tfrac{h}{2}\mathbf{e}_k)
=h\,\partial_ku+\frac{h^{3}}{24}\partial_k^{3}u+O(h^{5}),
\end{equation}
with all derivatives evaluated at $\mathbf{m}$; the even-order terms cancel.
Squaring,
\begin{equation}
\bigl(u_i-u_j\bigr)^2
=h^{2}(\partial_ku)^{2}+\frac{h^{4}}{12}\,\partial_ku\,\partial_k^{3}u+O(h^{6}),
\end{equation}
so the contribution of direction $k$ to $\Ecal_h$ is
\begin{equation}
\frac{h^{d-2}}{2}\sum_{\mathbf{m}}
\left[h^{2}(\partial_ku)^{2}+\frac{h^{4}}{12}\partial_ku\,\partial_k^{3}u\right]
+O(h^{4})
= h^{d}\sum_{\mathbf{m}}
\left[\frac{(\partial_ku)^{2}}{2}+\frac{h^{2}}{24}\partial_ku\,\partial_k^{3}u\right]
+O(h^{4}).
\end{equation}
The summed Taylor remainder is $O(h^4)$ by rapid decay in case (i) and by
uniformity of the $C^5$ remainder on the compact torus in case (ii).
Write
$f_k=\tfrac12(\partial_ku)^{2}$ and
$g_k=\tfrac1{24}\partial_ku\,\partial_k^{3}u$.
In case (i), both $f_k$ and $g_k$ are Schwartz, so
Lemma~\ref{lem:quadrature} gives
$h^d\sum f_k=\int f_k+O(h^4)$ and
$h^2h^d\sum g_k=h^2\int g_k+O(h^4)$.
In case (ii), $u\in C^5$ implies $f_k\in C^4$ and $g_k\in C^2$, and the
same two estimates follow from Lemma~\ref{lem:quadrature}. Hence
\begin{equation}
\Ecal_h[u]=\sum_k\int\left[\frac{(\partial_ku)^{2}}{2}
+\frac{h^{2}}{24}\,\partial_ku\,\partial_k^{3}u\right]\dr+O(h^{4}).
\end{equation}
Integrating by parts, with boundary terms vanishing by rapid decay or
periodicity, yields
$\int\partial_ku\,\partial_k^{3}u\dr=-\int(\partial_k^{2}u)^{2}\dr$, which
gives \eqref{eq:continuum}.
\end{proof}

\begin{corollary}
Under the hypotheses of Theorem~\ref{thm:continuum} with $A[u]>0$,
$\bigl|\Ecal_h[u]-\tfrac12\int|\nabla u|^2\bigr|
=h^2\bigl(A[u]/24+O(h^2)\bigr)$.
\end{corollary}

The coefficient $-\tfrac1{24}$ agrees with the quartic term in the standard
nearest-neighbor Fourier symbol $2(1-\cos qh)/h^2$.

\begin{remark}
\label{rem:hypotheses}
Two failures of the hypotheses matter in practice, and both are tested in
Section~\ref{sec:results}. Appendix~\ref{app:continuum} summarizes where these
hypotheses enter the proof.
(i) Finite domains. On a box with Dirichlet conditions the quadrature
corrections need not vanish, and boundary terms survive integration by parts.
In one dimension the two cancel (Lemma~\ref{lem:interval}). They also vanish
individually when $u''=0$ at the walls, as for the particle-in-a-box
eigenfunctions, so \eqref{eq:continuum} holds unmodified, as
Table~\ref{tab:errconst} confirms.
(ii) Nuclear cusps. As a function on $\mathbb{R}^3$, a Coulomb-bound
amplitude such as hydrogen $1s$ is not differentiable at the nucleus and
therefore is not $C^{5}$, so Theorem~\ref{thm:continuum} does not strictly
apply. For the hydrogenic cusp, however, the $1/r$ singularity in the second
derivatives is locally square-integrable in three dimensions, so $A[u]$
remains finite. This is consistent with, but does not by itself prove, the
approximately second-order numerical convergence. Empirically the measured
error still tracks $-\tfrac{h^2}{24}A[u]$, with the ratio consistent with a
$1+O(h)$ approach rather than the $1+O(h^2)$ behavior of the smooth case
(Table~\ref{tab:errconst}); the observed convergence order remains
$p\approx2$. A local scaling argument suggests an $O(h^3)$ correction for
the hydrogenic cusp. Edges near the nucleus contribute at this order, and
the corresponding coefficient can depend on the position of the nucleus
within the grid cell. This is consistent with the observed $1+O(h)$
approach of the measured-to-leading-term ratio.
\end{remark}

For the three analytic benchmarks the constant is available in closed form:
\begin{equation}
\label{eq:Avalues}
A_{\text{HO}}=\frac34,
\qquad
A_{\text{box}}=\pi^{4},
\qquad
A_{1s}=3 ,
\end{equation}
for $u=\pi^{-1/4}e^{-x^2/2}$ on $\mathbb{R}$, $u=\sqrt{2}\sin(\pi x)$ on
$[0,1]$, and $u=\pi^{-1/2}e^{-r}$ on $\mathbb{R}^3$ respectively; the
derivations are given in Appendix~\ref{app:continuum}.

\section{Identification with the von Weizs\"acker functional}
\label{sec:identification}

Setting $u=\sqrt\rho$ in the continuum limit of Theorem~\ref{thm:continuum},
\begin{equation}
\Ecal[u]\equiv\frac12\int|\nabla u|^2\dr
=\frac12\int\bigl|\nabla\sqrt\rho\bigr|^2\dr
=\frac18\int\frac{|\nabla\rho|^2}{\rho}\dr .
\end{equation}
Because $[\rho]=L^{-d}$ and $[u]=L^{-d/2}$ for an $N$-electron density,
$[\Ecal_h]=L^{-2}$ in every spatial dimension. Multiplication by the kinetic
scale $\hbar^2/m$, whose dimensions are energy times length squared, therefore
produces an energy.
The lattice functional uses the dimensionless calibration $C=1$.%
\label{sec:shape-scale}
If $C$ is retained, the limit is $C\Ecal[\sqrt\rho]$ and the corresponding physical
functional is $C\TW[\rho]$. Agreement with the conventional von Weizs\"acker
normalization requires $C=1$. Restoring physical units through $\hbar^2/m$
defines the $J$-derived continuum functional
\begin{equation}
\label{eq:TJ}
T_{J}[\rho]\;=\;\frac{\hbar^{2}}{m}\,\Ecal\bigl[\sqrt\rho\bigr]
\;=\;\frac{\hbar^{2}}{2m}\int\bigl|\nabla\sqrt\rho\bigr|^{2}\dr .
\end{equation}

\begin{corollary}
\label{cor:TW}
\begin{equation}
\boxed{\;
T_{J}[\rho]=\TW[\rho]=\frac{\hbar^{2}}{8m}\,\IF[\rho] . \;}
\end{equation}
\end{corollary}

With the calibrated geometric-mean pair weight, this continuum limit is the
von Weizs\"acker functional, equivalently the Fisher information of the
density in energy units. In atomic units $T_J[\rho]=\Ecal[\sqrt\rho]$, which
is the form used in the numerical calculations.

Constrained variation of the resulting amplitude functional gives the usual
Laplacian kinetic operator. The derivation and the equivalent density Euler
equation are given in Appendix~\ref{app:variational}. At the discrete level,
Proposition~\ref{prop:gradient} gives the corresponding lattice Laplacian.

\section{Scope of the amplitude functional}
\label{sec:scope}

The pair-weighted lattice energy depends only on a real nonnegative
amplitude. For any $N$-representable density, \eqref{eq:split} decomposes
$\Ts=\TW+\Tth$ with $\Tth\ge0$, and the continuum limit of
Section~\ref{sec:identification} recovers only $\TW$.

Write a general one-particle wavefunction in Madelung
form~\cite{Madelung1927}, $\psi=\sqrt\rho\,e^{iS}$, where $S$ is a
dimensionless phase. Almost everywhere on the set where $\rho>0$,
\begin{equation}
\label{eq:madelung}
|\nabla\psi|^{2}
=\bigl|\nabla\sqrt\rho\bigr|^{2}+\rho\,|\nabla S|^{2},
\end{equation}
and the exact kinetic energy is
$T=\tfrac{\hbar^2}{2m}\int|\nabla\psi|^2 = \TW[\rho]
+\tfrac{\hbar^2}{2m}\int\rho|\nabla S|^{2}$. The lattice energy \eqref{eq:Eh}
equals the first term only. It is built from a real nonnegative amplitude and
has no phase variable, so the current term is absent. Here $T$ is the kinetic
energy of the specified current-carrying state. The density alone does not
fix its value, so $T$ is distinct from the constrained-search functional
$\Ts[\rho]$.

A plane wave on a ring of circumference $L$,
$\psi(x)=L^{-1/2}e^{ikx}$ with $\rho=1/L$ constant and
$k=2\pi n/L$, $n\in\mathbb{Z}$, has $\nabla\rho\equiv0$. Then
$\TW[\rho]=0$ while $T=\hbar^{2}k^{2}/(2m)>0$ for $k\neq0$: a current-carrying
state can have vanishing von Weizs\"acker energy and nonzero kinetic energy.
The missing energy is the current contribution
$\tfrac{\hbar^2}{2m}\int\rho|\nabla S|^{2}$.

For a general noninteracting many-electron system the amplitude-gradient
functional equals the $\TW$ term only. For the uniform electron gas,
$\nabla\rho=0$ gives $\TW=0$ while the fermionic kinetic energy density is the
finite Thomas--Fermi value $\tfrac{3}{10}(3\pi^2)^{2/3}\rho^{5/3}$; the entire
kinetic energy resides in $\Tth$. When $\Ts[\rho]>0$ the recovered fraction
$f_W=\TW[\rho]/\Ts[\rho]$ lies in $[0,1]$, and the deficit $1-f_W$ is the Pauli
fraction omitted by the construction. A determinant-level analogue on
closed-shell Hartree--Fock densities is recorded in
Appendix~\ref{app:scope-numerics}; these determinant-level fractions need not
equal the exact constrained-search Pauli fraction.

\section{Computational methods}
\label{sec:methods}

All calculations use Hartree atomic units. Script order and package versions
are given in \texttt{REPRODUCIBILITY.md}; numerical parameters are given
in Appendix~\ref{app:computational}.

\subsection{Ratio-form and Dirichlet-form lattice energies}

Three algebraically equivalent evaluations are compared:
\begin{align}
\label{eq:impl-ratio}
\Ecal_h^{\text{ratio}}
&=h^{d-2}\sum_{\langle ij\rangle}\sqrt{\rho_i\rho_j}\;
  J\!\left(\sqrt{\frac{\rho_j}{\rho_i}}\right),\\
\label{eq:impl-stable}
\Ecal_h^{\text{stable}}
&=2h^{d-2}\sum_{\langle ij\rangle}\sqrt{\rho_i\rho_j}\;
\sinh^2\!\left[\frac14\bigl(\log\rho_j-\log\rho_i\bigr)\right],\\
\label{eq:impl-dir}
\Ecal_h^{\text{Dirichlet}}
&=\frac{h^{d-2}}{2}\sum_{\langle ij\rangle}
  \left(\sqrt{\rho_i}-\sqrt{\rho_j}\right)^{2}.
\end{align}
The first is the literal algebraic formula for the pair-weighted cost.
The second uses the stable identity
$J(x)=2\sinh^2[\tfrac12\log x]$ through the difference
$\log\rho_j-\log\rho_i$, without first forming $\rho_j/\rho_i$, and avoids
subtractive cancellation near $x=1$; both ratio forms require positive
amplitudes. The third is the continuous node extension of
Proposition~\ref{prop:nodes}. Comparison of the three forms assesses their
numerical implementation and floating-point behavior;
Proposition~\ref{prop:identity} establishes their exact algebraic
equivalence. Here $T_{J,h}$
denotes their common value $\Ecal_h[\sqrt\rho]$ in atomic units, and
$T_J^{\mathrm{extrap}}$ denotes its extrapolation to $h=0$.

\subsection{One-electron analytic densities}

Three systems with known kinetic energies test distinct features:
\begin{itemize}
\item One-dimensional harmonic oscillator, $\psi_0=\pi^{-1/4}e^{-x^2/2}$,
      $T=\tfrac14$: smooth, rapidly decaying, satisfies every hypothesis of
      Theorem~\ref{thm:continuum}.
\item Particle in a box, $\psi=\sqrt{2/L}\sin(\pi x/L)$,
      $T=\pi^2/(2L^2)$: tests the node extension, since $u=0$ at both walls.
\item Hydrogen $1s$, $\psi=\pi^{-1/2}e^{-r}$, $T=\tfrac12$: tests the
      nuclear cusp on a three-dimensional Cartesian grid offset by $h/2$ to
      avoid sampling $r=0$.
\end{itemize}
For the three-dimensional cases the box half-width $L$ and the spacing $h$ are
converged separately: $L$ is increased at fixed $h$ until the energy is
stationary, and only then is the $h$-refinement performed.

\subsection{Molecular electronic-structure calculations}

Electronic-structure calculations use PySCF~\cite{PySCF2018,PySCF2020}.

For H$_2^+$ the one-electron problem is solved variationally within the chosen
atomic-orbital (AO) basis by a single generalized eigenvalue problem using the
aug-cc-pV5Z
basis~\cite{Dunning1989,Kendall1992}; no SCF is required. Three quantities are
compared at each geometry: the analytic orbital expectation value
$T_{\text{orbital}}=\langle\phi|-\tfrac12\nabla^2|\phi\rangle$ from AO
integrals; $\TW$ evaluated as $\tfrac18\int|\nabla\rho|^2/\rho$ on an
atom-centered Becke quadrature grid~\cite{Becke1988}; and the Cartesian-lattice value $\Ecal_h$. As an external check on the orbital, total
energies are compared against high-accuracy reference values obtained from a
large uncontracted even-tempered basis, which reproduces the accepted results of
Bates \emph{et al.}~\cite{Bates1953} at $R=1$ and $R=2\,a_0$ to
$7\times10^{-8}\,\Ha$.

\section{Results}
\label{sec:results}

The numerical calculations address lattice convergence, the $h^2$ error
law, floating-point stability, and node handling. Equation~\eqref{eq:prop1}
is established algebraically; the numerical comparisons assess its
implementation and continuum approximation.

\subsection{Finite-ratio behavior and floating-point evaluation}
\label{sec:res-identity}

\begin{figure}[H]
\centering
\includegraphics[width=\textwidth]{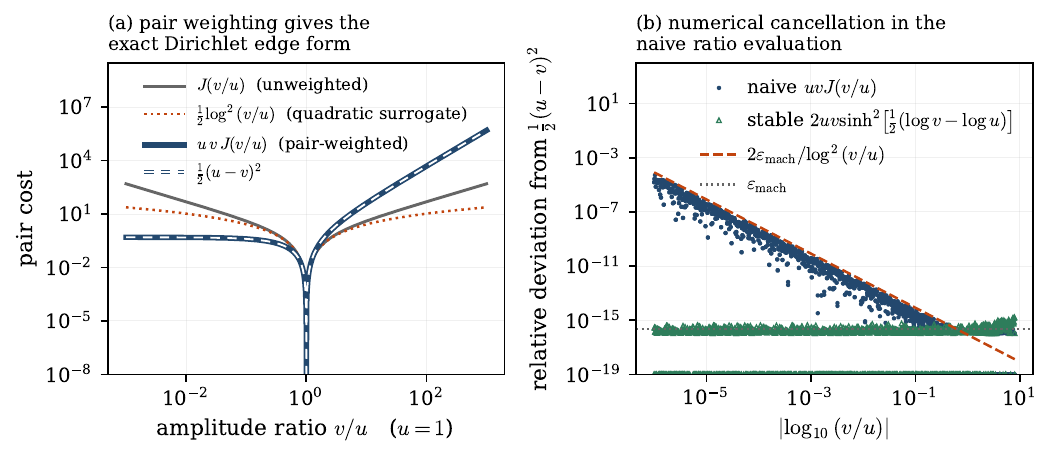}
\caption{\textbf{Unweighted cost, logarithmic surrogate, and pair-weighted cost.}
(a) The underlying $3201$-point sweep spans $10^{-8}\le v/u\le10^{8}$; the
panel displays the central six decades for readability. Over that range the
unweighted cost $J(v/u)$ (gray) and its near-equilibrium surrogate
$\tfrac12\log^2(v/u)$ (dotted) diverge from one another, while the
pair-weighted cost $uvJ(v/u)$ (solid blue) coincides with
$\tfrac12(u-v)^2$ (dashed overlay) everywhere.
(b) Relative deviations of the naive ratio evaluation
\eqref{eq:impl-ratio} and the stable log-ratio evaluation
\eqref{eq:impl-stable} from the Dirichlet form \eqref{eq:impl-dir}. Only the
naive formula exhibits the amplification
$2\varepsilon_{\text{mach}}/\log^{2}(v/u)$ derived in
\eqref{eq:cancellation}; the stable evaluation remains near the rounding
floor. The exact identity is algebraically regular; the loss of
floating-point accuracy arises from evaluating $J$ as
$\tfrac12(x+x^{-1})-1$ near $x=1$.}
\label{fig:identity}
\end{figure}

Near $v/u=1$, the naive ratio form loses precision
(Figure~\ref{fig:identity}b). The size of the loss
follows from \eqref{eq:Jcosh}: writing $x=v/u$ and $\epsilon=\log x$, the
quantity $\tfrac12(x+x^{-1})$ approaches $1$ from above like
$1+\tfrac12\epsilon^{2}$, so subtracting $1$ commits an absolute rounding
error of order $\varepsilon_{\text{mach}}$ in a result of size
$\tfrac12\epsilon^{2}$. The relative error therefore grows as
\begin{equation}
\label{eq:cancellation}
\frac{\varepsilon_{\text{mach}}}{\tfrac12\log^{2}(v/u)}
=\frac{2\varepsilon_{\text{mach}}}{\log^{2}(v/u)},
\end{equation}
which is the dashed reference curve in Figure~\ref{fig:identity}b. It
reproduces the measured scaling across six decades of $|\log_{10}(v/u)|$,
down to the point where the deviation reaches the $\varepsilon_{\text{mach}}$
floor, and overestimates its magnitude by a factor of about two.
At $v/u=1+10^{-8}$ the ratio form
returns exactly zero, a $100\%$ error, whereas the stable log-ratio formula
\eqref{eq:impl-stable} and the Dirichlet form agree to machine precision.
Thus the
cancellation is not unavoidable; the stable hyperbolic-sine representation
avoids the subtraction of nearly equal quantities. The Dirichlet
representation retains the additional practical advantage of remaining defined
at nodes.

At $R=2\,a_0$ and $h=0.25\,a_0$ on a separate identity-check box with
half-widths $L_{xy}=8\,a_0$ and $L_z=9\,a_0$ (not the production H$_2^+$
boxes of Appendix~\ref{app:computational}), the summed ratio and Dirichlet
energies agree to
$1.9\times10^{-16}$ relative despite an amplitude dynamic range of
$4.2\times10^{8}$, on $2.95\times10^{5}$ lattice points. This aggregate
agreement does not remove the local cancellation of the naive ratio formula
near $v/u=1$ identified in Figure~\ref{fig:identity}b.

On a finite hydrogen box at fixed $h=0.25\,a_0$, the pair-weighted energy
saturates at approximately $0.49183\,\Ha$ as the domain is enlarged;
subsequent mesh refinement gives $\TW=\tfrac12\,\Ha$. The unweighted cost
grows in proportion to the box volume
(Table~\ref{tab:unbridged}). In three dimensions this quantity has dimensions
of length, and its ratio to the predicted $|\Omega|/(2a_0^2)$ approaches
unity as the box is enlarged. The slow approach of that ratio to unity
reflects finite-box boundary-edge effects.

\begin{table}[H]
\centering
\small
\caption{\textbf{Pair-weighted energy and unweighted cost for hydrogen $1s$.}
Hydrogen $1s$ at $h=0.25\,a_0$ on cubic domains of half-width $L$. At this
fixed spacing, the pair-weighted energy approaches its infinite-domain
discrete value of approximately $0.49183\,\Ha$; mesh refinement separately
yields $\TW=0.5\,\Ha$. The unweighted cost of
Proposition~\ref{prop:unbridged} tracks $|\Omega|/(2a_0^2)$ and diverges
with the domain volume.}
\label{tab:unbridged}
\begin{tabularx}{\textwidth}{@{}Xrrrr@{}}
\toprule
$L$ ($a_0$) & $|\Omega|$ ($a_0^3$) & pair-weighted $\Ecal_h$ ($\Ha$)
& unweighted $\Scal_h^{\mathrm{unw}}$ ($a_0$)
& ratio to $|\Omega|/(2a_0^2)$\\
\midrule
3 & 216 & 0.475193 & 99.60 & 0.9222 \\
4 & 512 & 0.488881 & 241.27 & 0.9425 \\
6 & 1728 & 0.491750 & 831.71 & 0.9626 \\
8 & 4096 & 0.491826 & 1992.07 & 0.9727 \\
10 & 8000 & 0.491828 & 3914.89 & 0.9787 \\
\bottomrule

\end{tabularx}
\end{table}

\FloatBarrier
\subsection{Grid refinement and the leading
\texorpdfstring{$O(h^{2})$}{O(h squared)} term}
\label{sec:res-convergence}

\begin{figure}[H]
\centering
\includegraphics[width=0.63\textwidth]{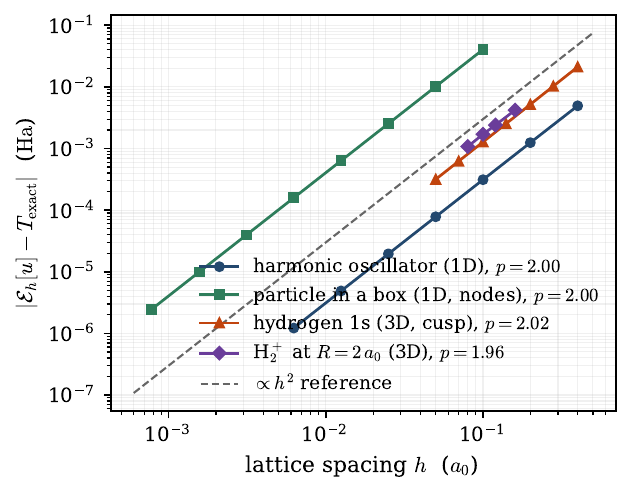}
\caption{\textbf{Lattice-to-continuum convergence.} Absolute error of
$\Ecal_h[u]$ against the exact kinetic energy, versus lattice spacing, for the
three analytic benchmarks and for H$_2^+$ at $R=2\,a_0$. Fitted orders $p$ are
given in the legend; the dashed line is an $h^2$ reference. The
particle-in-a-box series passes through nodes at both walls and converges at
the full second-order rate, confirming the node extension of
Proposition~\ref{prop:nodes} for these boundary nodes.}
\label{fig:convergence}
\end{figure}

All four benchmarks show approximately second-order convergence
(Figure~\ref{fig:convergence} and Table~\ref{tab:benchmarks}), with fitted
orders $p=2.00$, $2.00$, $2.02$ and $1.96$ for the harmonic oscillator, the
particle in a box, hydrogen $1s$ and H$_2^+$ at $R=2\,a_0$. The smooth
harmonic-oscillator result agrees directly with Theorem~\ref{thm:continuum};
the boundary-node and cusp cases, treated separately in
Remark~\ref{rem:hypotheses}, empirically retain second-order convergence.
Richardson extrapolation of the two finest spacings recovers the exact energies
to relative errors of $6\times10^{-11}$, $4\times10^{-13}$ and
$6\times10^{-6}$ for the three analytic systems.

\begin{table}[H]
\centering
\small
\caption{\textbf{One-electron benchmarks.} $T_{J,h}$ is the raw lattice value at
the finest spacing; $T_J^{\text{extrap}}$ is Richardson-extrapolated for the
analytic systems and least-squares extrapolated using the model
$T(h)=T_0+ah^2+bh^4$ for H$_2^+$. All energies in $\Ha$.}
\label{tab:benchmarks}
\begin{tabularx}{\textwidth}{@{}Xrrrcrr@{}}
\toprule
System & $T_{\text{ref}}$ & $T_{J,h}$ & rel.\ err. & $p$
& $T_J^{\text{extrap}}$ & rel.\ err.\\
\midrule
Harmonic oscillator (1-D) & 0.250000 & 0.249999 & $-4.9\times10^{-6}$ & 2.00 & 0.250000 & $-6.4\times10^{-11}$ \\
Particle in a box (1-D, nodes) & 4.934802 & 4.934800 & $-5.0\times10^{-7}$ & 2.00 & 4.934802 & $-4.0\times10^{-13}$ \\
Hydrogen $1s$ (3-D, cusp) & 0.500000 & 0.499684 & $-6.3\times10^{-4}$ & 2.02 & 0.500003 & $5.7\times10^{-6}$ \\
H$_2^+$ $1s\sigma_g$, $R=2\,a_0$ (3-D) & 0.602026 & 0.600949 & $-1.8\times10^{-3}$ & 1.96 & 0.602037 & $1.8\times10^{-5}$ \\
\bottomrule

\end{tabularx}
\end{table}

The fitted order alone does not distinguish among second-order
discretizations. The signed leading coefficient is recovered for the smooth
harmonic oscillator and for the boundary-compatible particle-in-a-box case:
the ratio of measured error to the leading $h^2$ term differs from unity by
$5.2\times10^{-5}$ at $h=0.025$ for the oscillator and by $1.3\times10^{-5}$
at $h=0.00625$ for the particle in a box (Table~\ref{tab:errconst}).
For hydrogen $1s$, which as a function on $\mathbb{R}^3$ is not
differentiable at the nucleus and therefore is not $C^5$, the ratio of
measured error to the leading $h^2$ term on the displayed spacings decreases from
$1.03057$ at $h=0.14$ to $1.01232$ at $h=0.05$. The approach is consistent
with $1+O(h)$ rather than $1+O(h^2)$, the degradation anticipated in
Remark~\ref{rem:hypotheses}.

\begin{table}[H]
\centering
\footnotesize
\setlength{\tabcolsep}{4pt}
\caption{\textbf{Comparison with the leading error coefficient.}
Measured error is $\Ecal_h[u]-T_{\mathrm{exact}}$. The leading $h^2$ term
is $-\tfrac{h^2}{24}A[u]$ with $A[u]$ evaluated in closed
form (Appendix~\ref{app:continuum}). The harmonic oscillator satisfies
Theorem~\ref{thm:continuum} directly; the particle-in-a-box case is covered
by Lemma~\ref{lem:interval}; hydrogen $1s$
violates the smoothness hypothesis because of the nuclear cusp and is
included as a diagnostic case. For hydrogen $1s$, this is a smooth-case
reference term rather than a theorem-backed error prediction. Four
representative spacings are shown for each system; the production finest
spacings used for the benchmark energies in Table~\ref{tab:benchmarks} are
listed in
Appendix~\ref{app:computational}.}
\label{tab:errconst}
\begin{tabularx}{\textwidth}{@{}Xccrrc@{}}
\toprule
System & $A[u]$ & $h$ & measured error & leading $h^2$ term & meas./lead.\\
\midrule
Harmonic oscillator & $3/4$ & 0.20000 & $-1.2458437\times10^{-3}$ & $-1.2500000\times10^{-3}$ & 0.99667 \\
 &  & 0.10000 & $-3.1223975\times10^{-4}$ & $-3.1250000\times10^{-4}$ & 0.99917 \\
 &  & 0.05000 & $-7.8108727\times10^{-5}$ & $-7.8125000\times10^{-5}$ & 0.99979 \\
 &  & 0.02500 & $-1.9530233\times10^{-5}$ & $-1.9531250\times10^{-5}$ & 0.99995 \\
\addlinespace
Particle in a box & $\pi^4$ & 0.05000 & $-1.0138439\times10^{-2}$ & $-1.0146780\times10^{-2}$ & 0.99918 \\
 &  & 0.02500 & $-2.5361735\times10^{-3}$ & $-2.5366951\times10^{-3}$ & 0.99979 \\
 &  & 0.01250 & $-6.3414117\times10^{-4}$ & $-6.3417377\times10^{-4}$ & 0.99995 \\
 &  & 0.00625 & $-1.5854141\times10^{-4}$ & $-1.5854344\times10^{-4}$ & 0.99999 \\
\addlinespace
Hydrogen $1s$ (cusp) & $3$ & 0.14000 & $-2.5248957\times10^{-3}$ & $-2.4500000\times10^{-3}$ & 1.03057 \\
 &  & 0.10000 & $-1.2788522\times10^{-3}$ & $-1.2500000\times10^{-3}$ & 1.02308 \\
 &  & 0.07000 & $-6.2279673\times10^{-4}$ & $-6.1250000\times10^{-4}$ & 1.01681 \\
 &  & 0.05000 & $-3.1634962\times10^{-4}$ & $-3.1250000\times10^{-4}$ & 1.01232 \\
\bottomrule

\end{tabularx}
\end{table}

\FloatBarrier
\subsection{\texorpdfstring{$\mathrm{H}_{2}^{+}$}{H2+} across internuclear separation}
\label{sec:res-h2plus}

\begin{figure}[H]
\centering
\includegraphics[width=\textwidth]{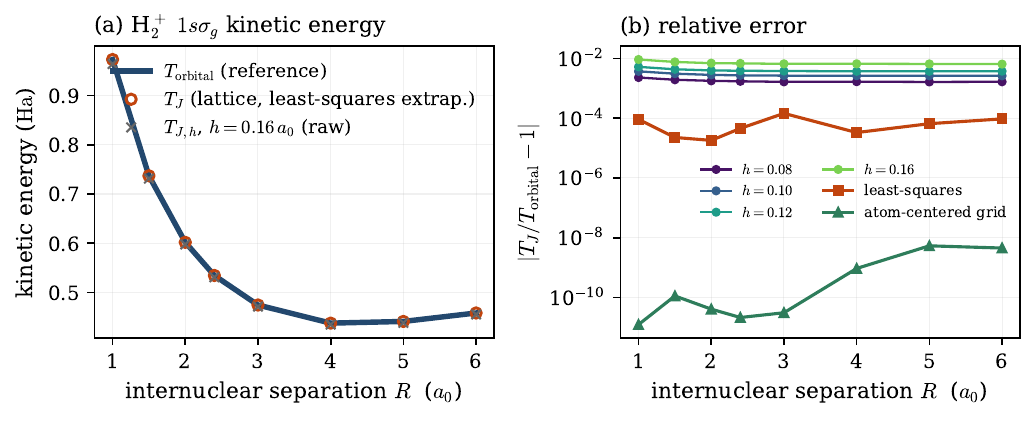}
\caption{\textbf{H$_2^+$ across the bond.}
(a) Kinetic energy of the $1s\sigma_g$ state versus internuclear separation.
The conventional orbital expectation value (line), the raw lattice
value at $h=0.16\,a_0$ (crosses) and the least-squares-extrapolated lattice
value (circles) are visually indistinguishable on this scale, including through the
non-monotonic region near $R=4\,a_0$.
(b) Relative error at each spacing. The atom-centered evaluation of
$\tfrac18\int|\nabla\rho|^2/\rho$ agrees with $T_{\text{orbital}}$ to
$\approx5\times10^{-9}$ relative, providing a lattice-independent numerical check of
the density/orbital identity in Corollary~\ref{cor:TW}.}
\label{fig:h2plus}
\end{figure}

H$_2^+$ is a molecular one-electron benchmark. Its real nodeless
$1s\sigma_g$ orbital makes $\TW$ the exact kinetic
energy rather than an approximation to it. The density changes shape
substantially across the scan, from a strongly bonded density at
$R=1\,a_0$ to two well-separated atomic lobes at $R=6\,a_0$.

Across $R=1$--$6\,a_0$, the atom-centered $\TW$ values reproduce the orbital
expectation to $\approx5\times10^{-9}$ relative, while the Cartesian-lattice values
converge to the same result with extrapolated relative errors below
$1.5\times10^{-4}$ (Figure~\ref{fig:h2plus} and Table~\ref{tab:h2plus}).
The atom-centered comparison checks the density quadrature against the
orbital expectation, with no Cartesian lattice involved; it is not an
independent prediction because both values use the same converged orbital.
The lattice comparison measures the discrete-to-continuum mapping and its
numerical error. Raw relative errors at $h=0.08\,a_0$ are uniformly
$\sim\!2\times10^{-3}$.

As an external check on the underlying orbital, the total energies
$E_{\text{el}}+1/R$, where $E_{\text{el}}$ is the electronic eigenvalue
excluding internuclear repulsion, obtained in the aug-cc-pV5Z basis agree with the
high-accuracy reference values to
between $1.1\times10^{-5}$ and $3.4\times10^{-5}\,\Ha$ over the whole scan,
consistent with basis-set incompleteness. The production $h^2+h^4$ fit has a
maximum relative deviation of $1.5\times10^{-4}$. Alternative extrapolation
models give deviations of up to $\sim5\times10^{-4}$, which we take as a
conservative sensitivity scale (Table~\ref{tab:extrap-robust}).

\begin{table}[H]
\centering
\small
\caption{\textbf{H$_2^+$ $1s\sigma_g$ kinetic energy across the bond.}
$T_{\text{orbital}}$ is the AO expectation value; $T_W^{\text{grid}}$ is
$\tfrac18\int|\nabla\rho|^2/\rho$ on an atom-centered Becke grid; $T_{J,h}$ is
the Cartesian lattice functional at $h=0.08\,a_0$; $T_J^{\text{extrap}}$ is
the least-squares extrapolation to $h\to0$. All energies in $\Ha$.}
\label{tab:h2plus}
\begin{tabularx}{\textwidth}{@{}Xrrcrcrc@{}}
\toprule
$R$ ($a_0$) & $T_{\text{orbital}}$ & $T_W^{\text{grid}}$ & difference
& $T_{J,h}$ & rel.\ err. & $T_J^{\text{extrap}}$ & rel.\ err.\\
\midrule
1.0 & 0.972684 & 0.972684 & $-1.2\times10^{-11}$ & 0.970436 & $-2.3\times10^{-3}$ & 0.972774 & $9.2\times10^{-5}$ \\
1.5 & 0.736929 & 0.736929 & $-8.2\times10^{-11}$ & 0.735495 & $-1.9\times10^{-3}$ & 0.736946 & $2.3\times10^{-5}$ \\
2.0 & 0.602026 & 0.602026 & $-2.4\times10^{-11}$ & 0.600949 & $-1.8\times10^{-3}$ & 0.602037 & $1.8\times10^{-5}$ \\
2.4 & 0.534825 & 0.534825 & $-1.1\times10^{-11}$ & 0.533902 & $-1.7\times10^{-3}$ & 0.534850 & $4.6\times10^{-5}$ \\
3.0 & 0.474872 & 0.474872 & $-1.5\times10^{-11}$ & 0.474085 & $-1.7\times10^{-3}$ & 0.474942 & $1.5\times10^{-4}$ \\
4.0 & 0.438154 & 0.438154 & $-4.1\times10^{-10}$ & 0.437420 & $-1.7\times10^{-3}$ & 0.438140 & $-3.3\times10^{-5}$ \\
5.0 & 0.441577 & 0.441577 & $-2.4\times10^{-9}$ & 0.440851 & $-1.6\times10^{-3}$ & 0.441606 & $6.6\times10^{-5}$ \\
6.0 & 0.458739 & 0.458739 & $-2.1\times10^{-9}$ & 0.457975 & $-1.7\times10^{-3}$ & 0.458782 & $9.5\times10^{-5}$ \\
\bottomrule

\end{tabularx}
\end{table}

\FloatBarrier
\section{Discussion}
\label{sec:discussion}

\subsection{Relation to Dirichlet and Fisher-information structures}

The von Weizs\"acker functional, its Fisher-information representation, and
the Dirichlet--Laplacian correspondence are established
structures~\cite{Weizsaecker1935,SearsParrDinur1980,Nagy2003,Nagy2025}.
For a probability vector on a path, Chen, Liu and Saloff-Coste define a
discrete Fisher information
$2\sum_i\bigl|\sqrt{p_i}-\sqrt{p_{i-1}}\bigr|^2$ and show that it equals a
multiple of the Dirichlet form of $\sqrt{p}$~\cite[p.~246]{ChenLiuSaloffCoste2008}.
The factorization $uvJ(v/u)=\tfrac12(u-v)^2$, together with the conditional
derivation of the $uv$ weight, maps the previously characterized ratio cost
exactly to that square-root Dirichlet form.
The same square-root coordinates underlie Hellinger and Fisher--Rao
geometry, so the edge form has the standard square-root Euclidean
structure~\cite{Holbrook2020}.

The composition law \eqref{eq:rcl}, as classified by
Washburn and Zlatanovi\'c~\cite{WashburnZlatanovic2026}, selects $J$
independently of this geometry. Assumption~\ref{ass:scale-deviation} and
degree-two covariance select the geometric-mean pair weight.
Proposition~\ref{prop:identity} converts the weighted
cost, at an arbitrary positive ratio, into a Dirichlet edge energy in the
amplitude.

\subsection{Comparison with the unweighted cost}

The unweighted and pair-weighted constructions differ both before and after
the continuum limit. The unweighted cost is only locally quadratic in the
logarithmic ratio and converges to the scale-invariant log-Dirichlet action
of Proposition~\ref{prop:unbridged}. In contrast,
$uvJ(v/u)=\tfrac12(u-v)^2$ holds at arbitrary positive ratio, and its lattice
limit is the von Weizs\"acker Dirichlet form.

\subsection{Scope and limitations}

More generally, Lemma~\ref{lem:structure} writes a general symmetric degree-two weight as
$B(u,v)=uv\,\chi(v/u)$ with $\chi(x)=\chi(1/x)$. Together with
Proposition~\ref{prop:identity},
\begin{equation*}
B(u,v)\,J\!\left(\frac{v}{u}\right)
=\chi\!\left(\frac{v}{u}\right)\cdot\frac12(u-v)^2 .
\end{equation*}
If $\chi$ is continuous at $x=1$, then $\chi(v/u)\to\chi(1)$ as
$v\to u$. For $u\in C^1(\mathbb{T}^d)$ with $u\ge u_{\min}>0$, the neighbor
ratios $u_j/u_i\to1$ uniformly as $h\to0$, so
$\sup_{\langle ij\rangle}\bigl|\chi(u_j/u_i)-\chi(1)\bigr|\to0$. The
Dirichlet edge sum $\tfrac{h^{d-2}}{2}\sum(u_i-u_j)^2$ remains bounded and
converges to $\tfrac12\int|\nabla u|^2$, hence
\begin{equation}
\label{eq:chi-limit}
\frac{h^{d-2}}{2}\sum_{\langle ij\rangle}
\chi\!\left(\frac{u_j}{u_i}\right)(u_i-u_j)^2
\;\xrightarrow[h\to0]{}\;
\frac{\chi(1)}{2}\int|\nabla u|^2\dr .
\end{equation}
Thus the leading continuum limit alone does not distinguish the
geometric-mean weight. Requiring
$B(u,v)J(v/u)=\tfrac{C}{2}(u-v)^2$ for arbitrary positive $u\neq v$ gives
$B(u,v)=Cuv$ off the diagonal. Since $J(1)=0$, exactness alone does not
determine $B(u,u)$; continuity of $B$ (equivalently of $\chi$ at $1$)
extends the result to the diagonal.

Theorem~\ref{thm:continuum} excludes generic nuclear cusps and finite domains
whose boundary terms do not cancel (Remark~\ref{rem:hypotheses}). Cartesian
lattices resolve the associated short core length scales inefficiently, as
the hydrogen calculations illustrate; adaptive or atom-centered
discretizations may improve molecular convergence. In general
$\TW\neq\Ts[\rho]$. Section~\ref{sec:scope} identifies the omitted current,
and Appendix~\ref{app:scope-numerics} records a closed-shell Hartree--Fock
survey of the determinant-level Pauli remainder. Open-shell and
current-carrying states remain outside the present numerical scope.

\section{Conclusion}
\label{sec:conclusion}

Under scale--deviation separation and degree-two covariance, the pair
weight has the form $B=Cuv$. With the dimensionless calibration $C=1$,
\begin{equation}
u\,v\,J\!\left(\frac{v}{u}\right)=\frac12(u-v)^{2}
\qquad\text{for all }u,v>0,
\end{equation}
so the pair-weighted $J$-cost is exactly a Dirichlet edge energy in amplitude space.
Retaining an arbitrary $C$ multiplies the lattice and
continuum functionals by $C$; the
choice $C=1$ matches the conventional von Weizs\"acker normalization. Under
$u=\sqrt\rho$, the calibrated lattice sum converges to the von Weizs\"acker
density-gradient form, with leading error
$-\tfrac{h^2}{24}\sum_k\int(\partial_k^2u)^2$.

The analytic, atomic, and molecular calculations test the lattice
implementation and its continuum scaling. The plane-wave identity and the
Hartree--Fock survey identify the omitted contributions as the current term
and the determinant-level Pauli remainder, respectively. Under these assumptions,
the reciprocal pair cost gives an exact discrete Dirichlet representation
whose continuum limit is the von Weizs\"acker sector, not a general
approximation to the full noninteracting kinetic functional $\Ts[\rho]$.

\section*{Funding}
\addcontentsline{toc}{section}{Funding}

This research received no external funding.

\section*{Conflicts of interest}
\addcontentsline{toc}{section}{Conflicts of interest}

The authors declare no conflict of interest.

\section*{Data and code availability}
\addcontentsline{toc}{section}{Data and code availability}

\begin{sloppypar}
\raggedright
The arXiv source package contains the manuscript source and generated table
and figure files. Ancillary files accompanying this preprint contain the
code, numerical results, and \texttt{REPRODUCIBILITY.md}. Package versions,
spacing sequences, and the role of each script are recorded in
\texttt{REPRODUCIBILITY.md}. Running \texttt{code/run\_all.py} executes
the calculations; \texttt{code/postprocess.py} and
\texttt{code/make\_figures.py} regenerate the tables and figures.
\end{sloppypar}

\appendix
\renewcommand{\thesection}{\Alph{section}}

\section{Continuum-limit details}
\label{app:continuum}

\subsection{Closed-form error constants}
\label{app:A}

We evaluate $A[u]=\sum_k\int(\partial_k^2u)^2\dr$ for the three analytic
benchmarks, giving the leading $h^2$ coefficients of
Table~\ref{tab:errconst}.

\paragraph{Harmonic oscillator.}
With $u=\pi^{-1/4}e^{-x^2/2}$ one has $u''=\pi^{-1/4}(x^2-1)e^{-x^2/2}$, so
\begin{equation}
A=\pi^{-1/2}\int_{-\infty}^{\infty}(x^2-1)^2e^{-x^2}dx
=\pi^{-1/2}\left(\tfrac34\sqrt\pi-2\cdot\tfrac12\sqrt\pi+\sqrt\pi\right)
=\frac34 .
\end{equation}

\paragraph{Particle in a box.}
With $u=\sqrt2\sin(\pi x)$ on $[0,1]$, $u''=-\sqrt2\pi^2\sin(\pi x)$ and
\begin{equation}
A=2\pi^{4}\int_0^{1}\sin^{2}(\pi x)\,dx=\pi^{4}\approx97.4091 .
\end{equation}
Note $u''(0)=u''(1)=0$, so the endpoint terms in
Lemma~\ref{lem:interval} vanish separately.

\paragraph{Hydrogen $1s$.}
With $u=\pi^{-1/2}e^{-r}=f(r)$, $f'=-f$ and $f''=f$. Writing $n_k=x_k/r$ and
$t_k=n_k^2$,
\begin{equation}
\partial_k^{2}u=f''n_k^{2}+\frac{f'}{r}\left(1-n_k^{2}\right)
=f\left[t_k-\frac{1-t_k}{r}\right],
\end{equation}
so
\begin{equation}
\sum_k\left(\partial_k^{2}u\right)^{2}
=f^{2}\left[\sum_kt_k^{2}
+\frac{2}{r}\sum_kt_k(t_k-1)
+\frac{1}{r^{2}}\sum_k(1-t_k)^{2}\right].
\end{equation}
Using the angular averages $\langle n_k^2\rangle=\tfrac13$ and
$\langle n_k^4\rangle=\tfrac15$ gives $\sum_k\langle t_k^2\rangle=\tfrac35$,
$\sum_k\langle t_k(t_k-1)\rangle=-\tfrac25$ and
$\sum_k\langle(1-t_k)^2\rangle=\tfrac85$. With $f^2=\pi^{-1}e^{-2r}$,
\begin{equation}
A=\int_0^{\infty}4\pi r^{2}\cdot\frac{e^{-2r}}{\pi}
\left[\frac35-\frac{4}{5r}+\frac{8}{5r^{2}}\right]dr
=4\left[\frac35\cdot\frac14-\frac45\cdot\frac14+\frac85\cdot\frac12\right]
=3 .
\end{equation}

\subsection{Quadrature remainders}
\label{app:quadrature}

\begin{lemma}
\label{lem:quadrature}
Let $h>0$ and let the sum run over a full Cartesian lattice of spacing $h$,
with any fixed shift.
\begin{enumerate}[label=(\roman*)]
\item If $f\in\mathcal{S}(\mathbb{R}^d)$, then
      $h^d\sum f=\int f+O(h^N)$ for every fixed integer $N\ge1$.
\item If $f\in C^4(\mathbb{T}^d)$ is sampled on a product grid commensurate
      with the periods, the product midpoint rule has error $O(h^4)$. If
      $g\in C^2(\mathbb{T}^d)$ on the same grid, the error is $O(h^2)$.
\end{enumerate}
\end{lemma}

\begin{proof}
(i) All derivatives of $f$ are integrable and decay rapidly. The
Euler--Maclaurin formula in each coordinate therefore contributes only
remainder terms: every boundary contribution at infinity vanishes, and the
remainder after $N$ derivatives is $O(h^N)$ for each fixed $N\ge1$.
(ii) On a circle, Taylor expansion about each cell midpoint gives
$\int_{\mathrm{cell}}\phi=h\phi(m)+\tfrac{h^3}{24}\phi''(m)+O(h^5)$
when $\phi\in C^4$, and $\int_{\mathrm{cell}}\phi=h\phi(m)+O(h^3)$ when
$\phi\in C^2$. Summing $O(h^{-1})$ cells, the $C^2$ case has error $O(h^2)$.
In the $C^4$ case the summed $\phi''$ correction is
$\tfrac{h^2}{24}\int\phi''+O(h^4)$, and $\int\phi''=0$ by periodicity,
leaving $O(h^4)$. The $d$-dimensional claim follows by iterating the
one-dimensional rule on the product grid.
\end{proof}

\begin{lemma}
\label{lem:interval}
Let $u\in C^5([0,L])$ be sampled at the vertices $x_i=ih$, $h=L/N$. Then
\begin{equation}
\label{eq:interval}
\Ecal_h[u]
=\frac12\int_0^L(u')^2\,dx
-\frac{h^2}{24}\int_0^L(u'')^2\,dx
+O(h^4).
\end{equation}
\end{lemma}

\begin{proof}
By \eqref{eq:latticeeq} with $d=1$,
$\Ecal_h[u]=\tfrac1{2h}\sum_{i=0}^{N-1}(u_{i+1}-u_i)^2$.
A symmetric Taylor expansion about each edge midpoint gives
$(u_{i+1}-u_i)^2=h^2(u')^2+\tfrac{h^4}{12}u'u'''+O(h^6)$, hence
\begin{equation}
\Ecal_h[u]
=h\sum_{i=0}^{N-1}\Bigl(\tfrac12(u')^2+\tfrac{h^2}{24}u'u'''\Bigr)+O(h^4),
\end{equation}
with derivatives at the midpoints. Cellwise Taylor expansion of an integrand
$G$ yields
$\int_{\mathrm{cell}}G=hG(m)+\tfrac{h^3}{24}G''(m)+O(h^5)$; summing and using
$h\sum G''=\int G''+O(h^2)=[G']_0^L+O(h^2)$ produces the composite midpoint
formula
$h\sum G=\int G-\tfrac{h^2}{24}[G']_0^L+O(h^4)$.
Applied to $G=\tfrac12(u')^2$, this is
\begin{equation}
h\sum\tfrac12(u')^2
=\tfrac12\int_0^L(u')^2\,dx
-\tfrac{h^2}{24}\bigl[u'u''\bigr]_0^L+O(h^4),
\end{equation}
while
\begin{equation}
h\sum u'u'''
=\int_0^L u'u'''\,dx+O(h^2)
=\bigl[u'u''\bigr]_0^L-\int_0^L(u'')^2\,dx+O(h^2).
\end{equation}
The endpoint terms cancel.
\end{proof}

\subsection{Hypotheses of the midpoint expansion}
\label{app:B}

Theorem~\ref{thm:continuum} is proved in Section~\ref{sec:continuum}. The
list below records where each hypothesis enters that proof, and how the
numerical calculations proceed when a hypothesis fails.

\begin{enumerate}
\item Schwartz regularity on $\mathbb{R}^d$, or $C^5$ regularity on a
      torus, controls the globally summed Taylor remainders and the
      quadrature orders of Lemma~\ref{lem:quadrature} in the stated
      $O(h^4)$ expansion. For the hydrogen $1s$ cusp studied
      here, the measured error ratio is empirically consistent with a
      $1+O(h)$ correction while the observed convergence order remains
      $p\approx2$.
\item Rapid decay or periodicity is used to identify the lattice sums with
      integrals at the orders of Lemma~\ref{lem:quadrature}, and to discard
      the boundary term in
      $\int\partial_ku\,\partial_k^{3}u=-\int(\partial_k^2u)^2$.
\item The edge-midpoint expansion is used so that even-order Taylor terms
      cancel in each edge difference. This cancellation follows by expanding
      about each edge's own midpoint and does not require a specially shifted
      vertex lattice. A finite interval in one dimension needs no extra
      boundary hypothesis: the endpoint corrections cancel
      (Lemma~\ref{lem:interval}). In higher dimensions
      the corresponding transverse quadrature terms need not cancel;
      conditions such as $\partial_k^2u=0$ on faces normal to $\mathbf{e}_k$
      provide convenient special cases.
\item Finite domains in the numerical work are handled by converging the
      box size at fixed $h$ before refining $h$. For hydrogen $1s$ the energy
      changes by $4\times10^{-8}\,\Ha$ between $L=10$ and $L=14\,a_0$, and by
      $9\times10^{-10}\,\Ha$ between $L=12$ and $L=14\,a_0$; the production
      value $L=14\,a_0$ is therefore converged well below the discretization
      error. For H$_2^+$ at $R=2\,a_0$ the energy changes by
      $2\times10^{-9}\,\Ha$ between $L_{xy}=10$ and $L_{xy}=14\,a_0$.
\end{enumerate}

\section{Variational details}
\label{app:variational}

\paragraph{Weighted graph Laplacian.}
For $\Ecal_G[u]=\tfrac12\sum_{(i,j)\in E}w_{ij}(u_i-u_j)^2$,
\begin{equation}
\frac{\partial\Ecal_G}{\partial u_i}=\sum_{j\sim i}w_{ij}(u_i-u_j)=(Lu)_i,
\end{equation}
with $L=D-W$ the weighted graph Laplacian, $D=\mathrm{diag}(\sum_jw_{ij})$.
Since $\Ecal_G[u]=\tfrac12u^{\top}Lu$ and $L\succeq0$, the energy is a
positive-semidefinite quadratic form whose kernel consists of the
locally constant fields.

\paragraph{Formal amplitude Euler equation.}
For sufficiently smooth $u$ and real $V$, with decay, periodic, or
fixed-Dirichlet boundary conditions that eliminate the boundary term in
integration by parts, consider
\begin{equation}
\mathcal{F}[u]=\frac{\hbar^{2}}{2m}\int|\nabla u|^{2}\dr+\int V(\rr)\,u^{2}\dr
\end{equation}
and the constraint $\int u^{2}\dr=N$. Introducing a chemical-potential
multiplier $\mu$ and using
$\delta\int|\nabla u|^2=-2\int(\nabla^2u)\,\delta u$ for admissible
variations, stationarity of $\mathcal{F}[u]-\mu\bigl(\int u^2-N\bigr)$ gives
\begin{equation}
\label{eq:schr}
-\frac{\hbar^{2}}{2m}\nabla^{2}u+Vu=\mu u ,
\end{equation}
so the conventional kinetic operator
$\hat T=-\tfrac{\hbar^{2}}{2m}\nabla^{2}$ is the Euler--Lagrange operator of
the $J$-derived amplitude-gradient functional.

For $N=1$, \eqref{eq:schr} is the ordinary normalized one-particle equation.
For general $N$, it is the Euler equation of the amplitude-gradient functional
$\TW+\int V\rho$ alone; it is not the complete interacting or many-orbital
Schr\"odinger problem. Expressed in the density, using
$\tfrac{\delta\TW}{\delta\rho}
=-\tfrac{\hbar^2}{2m}\tfrac{\nabla^2\sqrt\rho}{\sqrt\rho}$, the same condition
is the Euler equation of orbital-free
theory~\cite{LevyPerdewSahni1984}
\begin{equation}
-\frac{\hbar^{2}}{2m}\frac{\nabla^{2}\sqrt\rho}{\sqrt\rho}+V=\mu ,
\end{equation}
which holds pointwise where $\rho>0$ (and in the corresponding weak form
across zeros). It is exact for a one-orbital density and acquires the Pauli
potential $\delta\Tth/\delta\rho$ otherwise.

\section{Numerical details and sensitivity checks}
\label{app:computational}

\subsection{Numerical parameters}
\label{app:E}

\paragraph{Software versions.}
Python 3.11.4; NumPy 2.4.4; SciPy 1.17.1; SymPy 1.14.0; Matplotlib 3.11.1;
and PySCF 2.14.0. Exact package pins, commands, random seed, geometries and
complete spacing sequences are given in \texttt{REPRODUCIBILITY.md} in the
ancillary files.

\paragraph{Grids and domains.}
One-dimensional benchmarks use vertex grids including endpoints; the
particle-in-a-box well has width $L=1\,a_0$. The
three-dimensional benchmarks use grids offset by $h/2$ in every direction so
that no point coincides with a nucleus. Box half-widths: $L=12\,a_0$ (harmonic
oscillator), $L=14\,a_0$ (hydrogen), $L_{xy}=12\,a_0$ and
$L_z=12+R/2\,a_0$ (H$_2^+$ production grids), $L_{xy}=9\,a_0$ (He, H$_2$).
The floating-point identity check in Section~\ref{sec:res-identity} uses a
coarser separate box with $L_{xy}=8\,a_0$ and $L_z=9\,a_0$ at $h=0.25\,a_0$.
Finest spacings:
$h=6.25\times10^{-3}\,a_0$ (harmonic oscillator),
$7.81\times10^{-4}\,a_0$ (particle in a box), $0.05\,a_0$ (hydrogen,
$560^3=1.76\times10^{8}$ points), $0.08\,a_0$ (molecular).

\paragraph{Basis sets and quadrature.}
PySCF 2.14~\cite{PySCF2018,PySCF2020}; aug-cc-pV5Z for
H$_2^+$~\cite{Dunning1989,Kendall1992}, aug-cc-pVQZ for the many-electron set;
SCF convergence tolerance $10^{-12}$. Becke atom-centered
grids~\cite{Becke1988} at level 9 with pruning disabled; grid-level convergence
from level 7 to 9 is reported for every system in the result files.

\paragraph{Near-zero densities and particle number.}
Densities below $\rho_{\text{floor}}=10^{-14}$ are excluded from the
$|\nabla\rho|^2/\rho$ integrand. Electron-number integration on the production
(level 9) quadrature grids recovers $N$ to $\le3.9\times10^{-8}$ for all
many-electron systems and to $3.6\times10^{-12}$ for H$_2^+$. The ratio-form
implementation raises an error rather than returning a value if any amplitude
is nonpositive, which is how the particle-in-a-box node test confirms that the
ratio representation genuinely fails there while the extension does not.

\paragraph{Virial ratios.}
The virial ratio $-E_{\text{HF}}/\Tdet$, where $E_{\text{HF}}$ is the total
Hartree--Fock energy, lies within $7.0\times10^{-3}$ of unity for all nine
solutions and within $2.2\times10^{-3}$ for the other eight. The
H$_2$ value, $1.00696$, reflects in part the use of a near-experimental rather
than Hartree--Fock-optimized geometry; in the complete-basis limit at a
stationary geometry, the virial ratio is unity.

\begin{table}[H]
\centering
\small
\caption{\textbf{Density-floor sensitivity of $\TW$.} Level-9 unpruned Becke
grids and the same converged RHF density are used at each cutoff. The middle
column gives the production value at $\rho_{\mathrm{floor}}=10^{-14}$; the next
two give changes on using $10^{-12}$ and $10^{-16}$. Energies are in $\Ha$.}
\label{tab:floor}
\begin{tabularx}{\textwidth}{@{}Xrrrr@{}}
\toprule
System & $\TW(10^{-14})$ & $\Delta\TW(10^{-12})$
& $\Delta\TW(10^{-16})$ & relative spread\\
\midrule
He & 2.861170178 & $-3.7\times10^{-10}$ & $4.2\times10^{-12}$ & $1.3\times10^{-10}$ \\
H$_2$ & 1.125638782 & $-1.0\times10^{-8}$ & $7.1\times10^{-10}$ & $9.8\times10^{-9}$ \\
LiH & 7.697808085 & $-3.1\times10^{-9}$ & $8.3\times10^{-12}$ & $4.0\times10^{-10}$ \\
Be & 13.661972827 & $-4.7\times10^{-10}$ & $7.7\times10^{-12}$ & $3.5\times10^{-11}$ \\
Ne & 90.609333580 & $-4.4\times10^{-10}$ & $1.3\times10^{-11}$ & $5.0\times10^{-12}$ \\
CH$_4$ & 32.867422028 & $-1.4\times10^{-9}$ & $6.4\times10^{-14}$ & $4.3\times10^{-11}$ \\
H$_2$O & 57.613689310 & $-4.4\times10^{-10}$ & $2.3\times10^{-12}$ & $7.7\times10^{-12}$ \\
N$_2$ & 86.421384852 & $-7.3\times10^{-10}$ & $5.3\times10^{-12}$ & $8.5\times10^{-12}$ \\
CO & 88.354327618 & $-7.6\times10^{-10}$ & $9.2\times10^{-12}$ & $8.7\times10^{-12}$ \\
\bottomrule

\end{tabularx}
\end{table}

\noindent Across the nine systems the largest absolute spread is $1.1\times10^{-8}$\,$\mathrm{Ha}$ (H$_2$), a relative spread of $9.8\times10^{-9}$. The cutoff is therefore not a significant source of error at the level of the determinant Pauli fractions reported in Table~\ref{tab:manyelectron}.

\paragraph{Richardson and least-squares extrapolation.}
Analytic benchmarks use two-point Richardson extrapolation at order $2$. For
the three-dimensional molecular data, two-point Richardson is sensitive to the
sub-cell placement of nuclear peaks on an offset grid. The molecular energies are therefore also fitted by
least squares to $T(h)=T_0+ah^2+bh^4$ over the full refinement
sequence. That fit is the one reported in
Tables~\ref{tab:benchmarks} and~\ref{tab:h2plus} for H$_2^+$.

\begin{table}[H]
\centering
\small
\caption{\textbf{H$_2^+$ extrapolation robustness.} Relative deviations of
the extrapolated $T_0$ from $T_{\mathrm{orbital}}$. ``All'' uses
$h=0.16,0.12,0.10,0.08\,a_0$; ``drop'' omits $0.16\,a_0$; ``Rich.'' is
two-point Richardson from the two finest spacings. The three-point
$h^2+h^4$ fit has no residual degree of freedom and is included as a
sensitivity diagnostic, not as the production estimator.}
\label{tab:extrap-robust}
\begin{tabularx}{\textwidth}{@{}Xrrrrr@{}}
\toprule
$R$ ($a_0$) & $h^2+h^4$ all & $h^2$ all & $h^2+h^4$ drop
& $h^2$ drop & Rich.\\
\midrule
1.0 & $9.2\times10^{-5}$ & $-5.8\times10^{-5}$ & $5.5\times10^{-4}$ & $6.1\times10^{-6}$ & $2.0\times10^{-4}$ \\
1.5 & $2.3\times10^{-5}$ & $-8.3\times10^{-5}$ & $4.0\times10^{-4}$ & $-4.0\times10^{-5}$ & $1.1\times10^{-4}$ \\
2.0 & $1.8\times10^{-5}$ & $-1.0\times10^{-4}$ & $2.6\times10^{-4}$ & $-4.6\times10^{-5}$ & $6.1\times10^{-5}$ \\
2.4 & $4.6\times10^{-5}$ & $-4.4\times10^{-5}$ & $1.7\times10^{-4}$ & $-3.8\times10^{-7}$ & $5.9\times10^{-5}$ \\
3.0 & $1.5\times10^{-4}$ & $-5.2\times10^{-5}$ & $3.2\times10^{-4}$ & $4.8\times10^{-5}$ & $1.4\times10^{-4}$ \\
4.0 & $-3.3\times10^{-5}$ & $-3.3\times10^{-5}$ & $2.4\times10^{-4}$ & $-4.3\times10^{-5}$ & $5.8\times10^{-5}$ \\
5.0 & $6.6\times10^{-5}$ & $-4.0\times10^{-5}$ & $3.9\times10^{-4}$ & $4.8\times10^{-6}$ & $1.4\times10^{-4}$ \\
6.0 & $9.5\times10^{-5}$ & $-1.1\times10^{-4}$ & $1.6\times10^{-4}$ & $3.1\times10^{-7}$ & $5.8\times10^{-5}$ \\
\bottomrule

\end{tabularx}
\end{table}

\noindent Every estimator agrees with $T_{\text{orbital}}$ to within $5.5\times10^{-4}$ relative at every geometry: the production fit to $1.5\times10^{-4}$ and the two-parameter fit to $1.1\times10^{-4}$. The maximum least-squares residual among the four fitted lattice energies is $4.9\times10^{-5}$\,$\mathrm{Ha}$, and at any single geometry the five estimates span at most $5.9\times10^{-4}$\,$\mathrm{Ha}$.

\section{Reciprocal-cost characterizations}
\label{app:characterizations}

\subsection{Uniqueness of \texorpdfstring{$J$}{J}}
\label{app:uniqueness}

The composition law \eqref{eq:rcl} is a d'Alembert equation in logarithmic
coordinates. Its continuous solutions are classical~\cite{Aczel1966}.
The curvature condition itself implies continuity. The published
classification reaches the same conclusion from the composition law and
the unit-curvature calibration~\cite{WashburnZlatanovic2026}.

\begin{theorem}
\label{thm:Juniqueness}
Let $F:(0,\infty)\to\mathbb{R}$ satisfy the composition law
\eqref{eq:rcl} and have unit logarithmic curvature,
$\lim_{t\to0}2F(e^t)/t^2=1$. Then $F=J$.
\end{theorem}

\begin{proof}
Set $y=1$ in \eqref{eq:rcl}. Either $F(1)=0$, or $F(x)=-1$ for every
$x>0$; the latter is incompatible with the curvature condition. Thus
$F(1)=0$ and $H(t)=F(e^t)+1$ has $H(0)=1$. Substitution of $x=e^t$ and
$y=e^s$ gives
\begin{equation}
H(t+s)+H(t-s)=2H(t)H(s).
\end{equation}
Taking $t=0$ shows that $H$ is even. The curvature condition gives
$H(s)-1=\tfrac{s^2}{2}\cdot\tfrac{2F(e^s)}{s^2}\to0$, so $H(s)\to1$ as
$s\to0$. Setting $s=t$ gives $H(2t)=2H(t)^2-1$; applying the equation to
the pair $(t+s,t-s)$ then gives $H(t+s)H(t-s)=H(t)^2+H(s)^2-1$. Combined
with the equation itself,
\[
\bigl(H(t+s)-H(t-s)\bigr)^2=4\bigl(H(t)^2-1\bigr)\bigl(H(s)^2-1\bigr).
\]
As $s\to0$ the sum $H(t+s)+H(t-s)$ tends to $2H(t)$ and the difference
tends to $0$, so $H(t+s)\to H(t)$ and $H$ is continuous. The continuous
solutions with $H(0)=1$ are $\cosh\lambda t$ and $\cos\lambda t$ with
$\lambda\ge0$~\cite{Aczel1966}. For these, $2(H(t)-1)/t^2$ tends to
$\lambda^2$ and $-\lambda^2$ respectively, so the calibration selects
$H(t)=\cosh t$, and hence
$F(x)=\cosh(\log x)-1=\tfrac12(x+x^{-1})-1=J(x)$.
\end{proof}

\subsection{Quadratic edge characterization}
\label{app:rigidity}

Taking the edge cost to be a quadratic form in the amplitudes produces a
linear variational operator and a Dirichlet continuum form. The statements
below are converse characterizations within the geometric-mean pair-weight
class; they are not used in the forward derivation of
Proposition~\ref{prop:identity}.

\begin{theorem}
\label{thm:rigidity}
Let $\kappa>0$ and let $F:(0,\infty)\to[0,\infty)$. Consider the edge cost
\begin{equation}
\Kcal(u,v)=\kappa\,u\,v\,F\!\left(\frac{v}{u}\right),
\qquad u,v>0,
\end{equation}
and suppose that
\begin{enumerate}[label=(R\arabic*)]
\item $\Kcal$ is the restriction to the positive quadrant of a quadratic form
      $\Kcal(u,v)=\alpha u^2+\beta uv+\gamma v^2$;
\item $\Kcal$ vanishes at equilibrium, $\Kcal(u,u)=0$ for all $u>0$
      (equivalently $F(1)=0$).
\end{enumerate}
Then $\Kcal$ is symmetric and
\begin{equation}
\label{eq:rigid}
\Kcal(u,v)=\alpha\,(u-v)^2
\qquad\text{and}\qquad
F(x)=\frac{2\alpha}{\kappa}\,J(x),
\end{equation}
with $\alpha\ge0$. If $\Kcal$ is not identically zero, then $\alpha>0$. In
particular, the choice $\kappa=1$, $\alpha=\tfrac12$ recovers
$F=J$ and $\Kcal(u,v)=\tfrac12(u-v)^2$.
\end{theorem}

\begin{proof}
Vanishing on the diagonal (R2) gives $(\alpha+\beta+\gamma)u^2=0$, hence
$\beta=-\alpha-\gamma$. Evaluating the form at $u=1$, $v=x$,
\begin{equation}
\kappa\,x\,F(x)=\alpha+\beta x+\gamma x^2=(1-x)(\alpha-\gamma x)
=(\alpha-\gamma)(1-x)+\gamma(1-x)^2 .
\end{equation}
The left side is nonnegative. At $x=1-t$ and $x=1+t$ with $0<t<1$ this
reads $\pm(\alpha-\gamma)t+\gamma t^2\ge0$, so $|\alpha-\gamma|\le\gamma t$
for every such $t$, and therefore $\gamma=\alpha$. Then $\beta=-2\alpha$ and
\begin{equation}
\Kcal(u,v)=\alpha u^2-2\alpha uv+\alpha v^2=\alpha(u-v)^2 ,
\end{equation}
which is symmetric. Evaluating at $x=2$ gives $\alpha=2\kappa F(2)\ge0$, and
\begin{equation}
F(x)=\frac{\alpha}{\kappa}\cdot\frac{(1-x)^2}{x}.
\end{equation}
By the factorization \eqref{eq:Jfact}, $(1-x)^2/x=2J(x)$, so
$F=(2\alpha/\kappa)J$. If $\Kcal$ is not identically zero, then $\alpha>0$.
\end{proof}

Within the quadratic edge class, nonnegativity together with vanishing on
the diagonal forces symmetry. Without nonnegativity,
$\Kcal(u,v)=(u-v)(\alpha u-\gamma v)$ with $\alpha\ne\gamma$ satisfies (R1)
and (R2) but changes sign across the diagonal.

The first variation of the quadratic form in (R1) is linear in the
amplitudes, and the resulting edge form has a Dirichlet continuum limit.

\begin{corollary}
\label{cor:joint-rigidity}
Suppose that a pair weight satisfies (B2) and
Assumption~\ref{ass:scale-deviation}. If the edge cost
$\Kcal(u,v)=B(u,v)F(v/u)$ is a nonzero,
nonnegative quadratic form that vanishes on the diagonal, then
\begin{equation}
B(u,v)=Cuv,\qquad
F(x)=\frac{2\alpha}{C}\,J(x)
\end{equation}
for constants $C>0$ and $\alpha>0$, with
$\Kcal(u,v)=\alpha(u-v)^2$.
\end{corollary}

\begin{proof}
Proposition~\ref{prop:scaleonly} gives $B(u,v)=Cuv$. Theorem~\ref{thm:rigidity}
then applies with $\kappa=C$.
\end{proof}

\subsection{Inversion covariance of the geometric mean}
\label{app:inversion}

Assumption~\ref{ass:scale-deviation} takes the geometric mean as the pair
scale. An additional inversion-covariance condition singles out the same
scale.
Fix a reference amplitude $u_\ast>0$ and call $u\mapsto u_\ast^2/u$
amplitude inversion; in the logarithmic coordinates \eqref{eq:log-midpoint}
it reverses the signs of $\sigma$ and $\delta$, and it maps the ratio $v/u$
to $u/v$, leaving $J(v/u)$ unchanged.

\begin{proposition}
\label{prop:inversion}
\leavevmode
\begin{enumerate}[label=(\roman*)]
\item Let $s:(0,\infty)^2\to(0,\infty)$ be symmetric, satisfy
      $s(cu,cv)=c\,s(u,v)$ for $c>0$, and be covariant under inversion,
      $s(u_\ast^2/u,u_\ast^2/v)=u_\ast^2/s(u,v)$. Then $s(u,v)=\sqrt{uv}$.
\item Let $B:(0,\infty)^2\to(0,\infty)$ satisfy \textup{(B1)}, \textup{(B2)}
      and $B(u_\ast^2/u,u_\ast^2/v)\,B(u,v)=C^2u_\ast^4$ for a constant
      $C>0$. Then $B(u,v)=C\,uv$.
\end{enumerate}
\end{proposition}

\begin{proof}
(i) Homogeneity with $c=u_\ast^2/(uv)$ maps the pair $(v,u)$ to
$(u_\ast^2/u,u_\ast^2/v)$, so by symmetry
$s(u_\ast^2/u,u_\ast^2/v)=u_\ast^2\,s(u,v)/(uv)$. Equating this with
$u_\ast^2/s(u,v)$ gives $s(u,v)^2=uv$.
(ii) The same substitution with \textup{(B2)} gives
$B(u_\ast^2/u,u_\ast^2/v)=u_\ast^4B(u,v)/(uv)^2$, so
$\bigl(B(u,v)/(uv)\bigr)^2=C^2$ and $B=Cuv$ by positivity.
\end{proof}

Inversion covariance is an additional structural condition, not a
consequence of exchange symmetry alone.

\section{Determinant-level Pauli fractions for closed-shell RHF systems}
\label{app:scope-numerics}

The calculations in this appendix illustrate the determinant-level Pauli
contribution omitted by the amplitude functional. They are not tests of the
discrete-to-continuum mapping.

Electronic-structure calculations use PySCF~\cite{PySCF2018,PySCF2020}.
Converged restricted Hartree--Fock determinants in aug-cc-pVQZ give
$\Tdet=\operatorname{Tr}[\mathbf{D}\mathbf{T}]$, where $\mathbf{D}$ is the
converged one-particle density matrix and $\mathbf{T}$ the kinetic-energy
matrix in the atomic-orbital basis, and give $\TW$ from the same density on a
Becke grid. The self-consistent-field calculation and grid level are
converged for every system; the spread of $\TW$ across grid levels 7--9 does
not exceed $1.6\times10^{-7}\,\Ha$. Densities below $10^{-14}$ are excluded
from the $|\nabla\rho|^2/\rho$ integrand; a same-grid sensitivity scan at
$10^{-12}$, $10^{-14}$ and $10^{-16}$ is reported with
Table~\ref{tab:floor}.

An RHF determinant need not attain the constrained-search minimum defining
$\Ts[\rho]$ for its density. Write
\begin{equation}
\TPdet=\Tdet-\TW[\rho_{\mathrm{HF}}],
\qquad
\fWdet=\frac{\TW[\rho_{\mathrm{HF}}]}{\Tdet}.
\end{equation}
Nonnegativity of $\TPdet$ follows from the chain
$\TW[\rho_{\mathrm{HF}}]\le\Ts[\rho_{\mathrm{HF}}]\le\Tdet$: the first
inequality is the lower-bound property of $\TW$, and the second holds because
the RHF determinant is one admissible wavefunction in the constrained search
over states with density $\rho_{\mathrm{HF}}$. For a real determinant with
one occupied spatial orbital, $\fWdet=1$. These determinant-level quantities
equal $\Tth[\rho]$ and $f_W$ only when the reference determinant realizes
$\Ts[\rho]$.

\begin{figure}[H]
\centering
\includegraphics[width=0.72\textwidth]{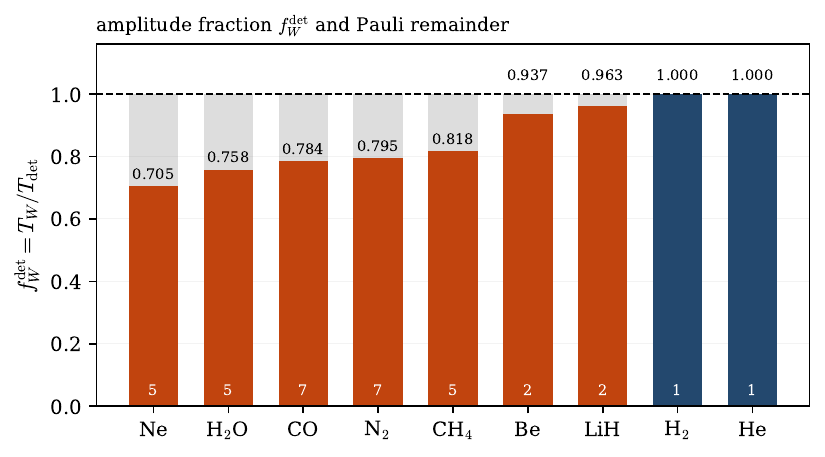}
\caption{\textbf{Determinant Pauli fractions $\fWdet$.}
$\fWdet=\TW/\Tdet$ for nine closed-shell RHF determinants; digits inside
the bars give the number of occupied spatial orbitals. The two
single-orbital systems (blue) are consistent with $\fWdet=1$ within
quadrature error; every multi-orbital system shown falls short, and the gray
remainder is the determinant-level Pauli term.}
\label{fig:scope}
\end{figure}

Table~\ref{tab:manyelectron} gives the decomposition. As expected from the
one-orbital limit, the production-grid values for He and H$_2$ differ from
unity only by numerical quadrature residuals: $1.6\times10^{-12}$ for He and
$7.0\times10^{-10}$ for H$_2$. These differences are not precision estimates;
changing the quadrature grid from level 7 to 9
changes $\TW$ by $1.6\times10^{-11}\,\Ha$ and
$2.8\times10^{-10}\,\Ha$, respectively.
On the same level-9 grid, varying the density floor from $10^{-12}$ to
$10^{-16}$ changes $\TW$ by at most $1.1\times10^{-8}\,\Ha$ among all nine
systems, or $9.8\times10^{-9}$ relative
(Table~\ref{tab:floor}). The cutoff is therefore negligible for the
reported Pauli fractions, although it is larger than some one-orbital
residuals and those residuals should not be read as precision estimates.
Every many-orbital system has
$\fWdet<1$, ranging from $0.963$ for
LiH down to $0.705$ for Ne. The deficit is not determined by orbital count
alone. The determinant-level Pauli fraction is $29.5\%$ for Ne and $18.2\%$
for CH$_4$.

The virial ratios $-E_{\text{HF}}/\Tdet$ lie near unity; the values and
the geometry caveat are given in Appendix~\ref{app:computational}.

\begin{table}[H]
\centering
\small
\caption{\textbf{Many-electron determinant decomposition
$\Tdet=\TW+\TPdet$.} Restricted
Hartree--Fock, aug-cc-pVQZ, near-experimental geometries. $N$ is the number of
electrons and $N_{\text{occ}}$ the number of occupied spatial orbitals. $\fWdet=\TW/\Tdet$ is unity in the
one-orbital limit; the deficit $1-\fWdet$ is the determinant-level Pauli
contribution. All energies in $\Ha$.}
\label{tab:manyelectron}
\begin{tabularx}{\textwidth}{@{}Xccrrrc@{}}
\toprule
System & $N$ & $N_{\text{occ}}$ & $\Tdet$ & $\TW$ & $\TPdet$ & $\fWdet$\\
\midrule
He & 2 & 1 & 2.86117 & 2.86117 & 0.00000 & 1.000000 \\
H$_2$ & 2 & 1 & 1.12564 & 1.12564 & 0.00000 & 1.000000 \\
LiH & 4 & 2 & 7.99037 & 7.69781 & 0.29256 & 0.963386 \\
Be & 4 & 2 & 14.57290 & 13.66197 & 0.91093 & 0.937492 \\
Ne & 10 & 5 & 128.53570 & 90.60933 & 37.92637 & 0.704935 \\
CH$_4$ & 10 & 5 & 40.18101 & 32.86742 & 7.31358 & 0.817984 \\
H$_2$O & 10 & 5 & 75.99146 & 57.61369 & 18.37777 & 0.758160 \\
N$_2$ & 14 & 7 & 108.76195 & 86.42138 & 22.34056 & 0.794592 \\
CO & 14 & 7 & 112.63365 & 88.35433 & 24.27932 & 0.784440 \\
\bottomrule

\end{tabularx}
\end{table}

\end{document}